\documentclass[12pt,a4paper]{article}
\pdfoutput=1

\usepackage[utf8]{inputenc}
\usepackage[T1]{fontenc}

\usepackage[margin=2.0 cm]{geometry}
\usepackage[pdfstartview={FitH},bookmarks=false,linktoc=page,colorlinks=true,linkcolor=blue,citecolor=blue,urlcolor=blue]{hyperref}
\usepackage[numbered]{bookmark}
\usepackage{mathtools}
\usepackage{amssymb,bm}
\usepackage{graphicx}
\usepackage[font=small,labelfont=bf]{caption}
\usepackage{authblk}
\usepackage{cite}
\usepackage{dsfont}
\usepackage[dvipsnames]{xcolor}
\usepackage[titles]{tocloft}
\usepackage{float}
\usepackage{caption}
\usepackage{subcaption}
\usepackage{color}
\usepackage{tikz}
\usepackage{ifthen}
\usepackage[warn]{textcomp}%for the number sign
\numberwithin{equation}{section}
\allowdisplaybreaks
\usepackage{multirow}

\usetikzlibrary{patterns}

\DeclareMathOperator{\tr}{Tr}

\usepackage{comment}

\newcommand{\overbar}[1]{\mkern 1.5mu\overline{\mkern-1.5mu#1\mkern-1.5mu}\mkern 1.5mu}
\begin{document}

%%%%%%%%%%%%%%%%%%%%%%%%%%%%%%%%%%%%%%%%%%%%
\title{\vspace{1.3cm}\textbf{From Heun to Painlev\'e on Sasaki-Einstein Spaces and Their Confluent Limits}\vspace{0.7cm}}
%%%%%%%%%%%%%%%%%%%%%%%%%%%%%%%%%%%%%%%%%%%%

\author[a]{V. Avramov}
\author[a,b]{H. Dimov}
\author[a]{M. Radomirov}
\author[a,c]{R. C. Rashkov}
\author[a]{T. Vetsov}

\affil[a]{\textit{Department of Physics, Sofia University,}\authorcr\textit{5 J. Bourchier Blvd., 1164 Sofia, Bulgaria}
	
	\vspace{-10pt}\texttt{}\vspace{0.0cm}}

\affil[b]{\textit{The Bogoliubov Laboratory of Theoretical Physics, JINR,}\authorcr\textit{141980 Dubna,
		Moscow region, Russia}
	
	\vspace{-10pt}\texttt{}\vspace{0.0cm}}

\affil[c]{\textit{
		Institute for Theoretical Physics, Vienna University of Technology,}\authorcr\textit{Wiedner Hauptstr. 8–10, 1040 Vienna, Austria}
	
	\vspace{10pt}\texttt{v.avramov,h\_dimov,radomirov,rash,vetsov@phys.uni-sofia.bg}\vspace{0.1cm}}

\date{}
\maketitle
\vspace{-20pt}

\begin{abstract}
The aim of this paper is to study the effect of isomonodromic deformations of the evolution of scalar fields in Sasaki-Einstein spaces in the context of holography. Here we analyze the monodromy data of the general Heun equation, resulting from a scalar on Y$^{p,q}$, thus obtaining the corresponding Painlev\'e VI equation. Furthermore we have considered limits leading to a coalescence of singularities, which in turn transform the original Painlev\'e VI equation, to one of lower rank. The confluent limits we have considered are Y$^{p,p}$, T$^{1,1} / \mathbb{Z}_2$ and Y$^{\infty, q}$.
\end{abstract}

\tableofcontents
\newpage

\section{Introduction}
One of the main characteristics of the AdS/CFT correspondence \cite{Maldacena:1997re} is the prediction of specific holographic dualities between gauge theories and string theory on certain geometric backgrounds. In particular, a duality between quiver gauge theories \cite{douglas1996d, Martelli:2004wu, Lechtenfeld:2014fza, Benvenuti:2004dy, Benvenuti:2005ja} and type IIB string theory on AdS$_5 \times$M$^5$ backgrounds has attracted great attention, with $M^5$ being an arbitrary compact Sasaki-Einstein manifold. This is motivated by the recent progress in the construction of such backgrounds \cite{Gauntlett:2004yd, Martelli:2005wy, Gauntlett:2004hh, Cvetic:2005ft, Cvetic:2005vk, Chen:2004nq, Lu:2005sn}. In this case, the Einstein property means that the Ricci tensor is proportional to the metric ${\rm Ric}_g=\lambda g$  for some constant $\lambda$, and Sasaki implies that the metric cone $\bar g=dr^2+r^2 g$ is K\"ahler. The main feature of the Sasaki-Einstein manifolds is that their restricted holonomy group of the cone is Hol$^{(0)}{\bar g}=SU(n)$, where $n$ is the complex dimension of the K\"ahler cone \cite{sparks2010sasaki}.

Sasaki-Einstein spaces inherit a number of geometric structures
from the K\"ahler structure of its cone. Specifically, a key role in the classification of the Sasaki-Einstein manifolds is played by
 the Reeb vector field defined as $\hat Q_R = J(r\partial r)$, where $J$ is an integrable complex structure on the K\"ahler cone. If all the orbits of $\hat Q_R$ are compact (i.e. circles) then the Sasakian manifold is said to be regular, otherwise it is quasi-regular. On the other hand, if $\hat Q_R$  has a non-compact orbit the Sasakian manifold is said to be irregular.  In the context of holography the Reeb vector is shown to correspond to the ${\mathcal R}$-symmetry of the dual gauge theory \cite{ashmore2016exceptional}.

Only recently particular examples of non-trivial irregular Sasaki-Einstein manifolds, namely the infinite families of Y$^{p,q}$ and L$^{p,q,r}$ spaces, have been discovered \cite{Gauntlett:2004yd, Cvetic:2005ft}. They have become an objects of significant interest in the context of holography due to the fact that their corresponding duals are quiver gauge theories \cite{Martelli:2004wu}. Usually quiver theories are super Yang-Mills theories whose field content is determined by a gauge group on each node of the quiver. The latter naturally appear in field theories coming from D-brane models in string theory with D-branes located at certain singularities of Calabi-Yau varieties \cite{douglas1996d}. The nodes  of the quiver for a ($d+1$)-dimensional quiver gauge theory correspond to a number of coincident D-branes and the edges of the quiver to open string states stretching between these branes \cite{lawrence1998conformal}.

An additional strong motivation for studying Sasaki-Einstein spaces comes from a recent investigation of their properties in the context of emergent spacetime \cite{berman2022emergent, berman2021emergent}. Here the classical geometry of the bulk spacetime is conjectured to emerge from a particular quantum state of the dual gauge theory. 

Furthermore, one of the fruitful areas of investigation of AdS/CFT comes from the powerful techniques of integrability. The latter allows for many new relations and structures to be uncovered, which  could significantly improve our understanding of the systems under consideration. In particular, recent advances in the theory of differential equations allowed for new and powerful analytical methods to appear. One such technique, namely the method of isomonodromic deformations \cite{jimbo1, jimbo2, jimbo3}, starts to draw attention, mostly because  it gives an alternative way of decoding physical information from the singularities of the modeling differential equations \cite{castro2013black, da2016kerr, novaes2014isomonodromy, da2015kerr, da2016existence, barragan2018scalar, da2020confluent, amado2020vector, da2021teukolsky, amado2021qnms, da2022expansions, bershtein2022quantum, jeong2020riemann, amado:2020, cavalcante:2019}. Hereby the method of isomonodromic deformations can provide an interesting approach for studying  integrable structures in holographic systems. One of the reasons we would like to consider this method in string theory is that its capabilities are not widely explored in the context of holography. 

 The main advantage of the isomonodromic method is that it allows one to obtain conditions for the change of monodromy data under which the original differential equation remains invariant and integrable. Furthermore, this method can be used to transform the boundary conditions of the original differential equation to a set of initial conditions for some nonlinear Painlev\'e type of equations and thus using the advancement in this area to extract relevant physical data \cite{amado:2020, cavalcante:2019, da2016kerr}. The latter conveys strong motivation to consider this techniques to study bulk field dynamics in Sasaki-Einstein spaces, which will be the focus of our investigation.

The starting point of the method of isomonodromic deformation is to consider a Fuchsian differential equation, which exhibits the Painlev\'e property. This means that the only movable singularities are simple poles. The essence of the method is finding non-trivial conditions, for which the monodromy data must belong to the equivalence class of the original differential equation. These conditions turn out to be integral curves generated by a specific Hamiltonian flow, whose equations of motion reduce to the family of non-linear Painlev\'e equations. Consequently, one can study the properties of the latter equations to gain insights for relevant structures of the original theory.

This work aims to study the evolution of scalar fields in Y$^{p,q}$ and some if its specific limits via the method of isomonodromic deformations. The main motivation is that if certain limits are considered in the general Y$^{p,q}$, the isometries of the geometry will change leading to simpler equations of motion, which by duality will also transform the corresponding superconformal theory to a simpler one. 
For example, let us comment on a few special cases, namely\footnote{It is worthy to note that $Y^{2,1}$ corresponds to del Pezzo$_1$ singularity. It gauge theory dual is also known, namely the quiver. The exact toric superpotential can be found in \cite{Feng:2000mi}.} $Y^{1,0}$, $Y^{2,2}$ and $Y^{1,1}$. The first one is a conifold while the others are related to weighted projective spaces. The case of the conifold
can be realized as a 3 (complex) dimensional manifold in $\mathbb{C}^4$ as a quadric
\begin{equation}
w_1^2+w_2^2+w_3^2+w_4^2=0.
\label{def-conifold--C4}
\end{equation}
The conical singularity is placed at $w_1=w_2=w_3=w_4=0$ and actually the manifold can be considered as a 5d real manifold $T^{1,1}$. The torus $T^{1,1}$ is $SU(2)\times SU(2)/U(1)$ with topology of $S^2\times S^3$. Another way to see it is as $U(1)$ fibration over $\mathbb{CP}^1\times\mathbb{CP}^1$, which is known to be K\"ahler-Einstein. In general the quivers have $2p$ gauge groups, thus the conifold has 2
gauge groups, and it corresponds to $Y^{1,0}$. Klebanov and Witten identified the potential of the corresponding 4d superconformal theory \cite{klebanov1998superconformal}
\begin{equation}
W_{KW}\propto \varepsilon^{\alpha\beta}\varepsilon^{\dot{\alpha}\dot{\beta}} \operatorname{Tr} \left(A_\alpha B_{\dot{\alpha}}A_\beta B_{\dot{\beta}}\right),
\label{KW-potential}
\end{equation}
where $(A_1,A_2)$ and $(B_1,B_2)$ are doublets of the preserved global $SU(2)\times SU(2)$ flavor symmetry. It is important to note that there is also a baryonic $U(1)$ symmetry which is not a consequence of the isometries of $T^{1,1}$, but rather due to $S^3$ cycle in the topology of $T^{1,1}$. According to the holographic correspondence the gauge theory barionic operator is dual to D3 branes wrapping supersymmetric 3-cycle.
This holographic correspondence was discovered first in \cite{klebanov1998superconformal} and triggered subsequent studies \cite{Gubser:1998vd, bah2014linear, bah2012ads} of quivers as holographic dual to certain geometries generalizing the above picture. The other members of the family $Y^{p,0}$ are just $\mathbb{Z}_p$ orbifolds of the conifold. In particular, $Y^{2,0}=Y^{1,0}/\mathbb{Z}_2$ where the $U(1)$ fiber over $\mathbb{CP}^1\times\mathbb{CP}^1$ has twice smaller length. We would also like to note that the cases $Y^{1,1}, Y^{1,0}$ and $Y^{2,2}$ are interesting as being the only members of $Y^{p,q}$ admitting massive supersymmetric deformations, see for instance \cite{Feng:2000mi}. 

The structure of the paper is the following. In Section \ref{ypqSect}, we show that the equation of motion for a scalar field on $Y^{p,q}$ reduces to a single general Heun equation with four regular singularities. In this case, the derived in Section \ref{minv} isomonodromy method, is trivially applicable and one can easily find the explicit form of the corresponding nonlinear Painleve VI equation. Consequently, we consider several special limits of the original geometry and study their properties. Namely, in the limit $q \to p$ the original Heun equation reduces to a hypergeometric one with resulting appearance of resonant singularities. Therefore the isomonodromic method is not directly applicable and should be further extended. In addition we look at the special case $Y^{1,0}$, which is known to be isomorphic to $T^{1,1} / \mathbb{Z}_2$, resulting in two hypergeometric equations, also having resonant singular points. Finally, we investigate the limit Y$^{\infty, q}$ and show that it reduces to a confluent Heun differential equation. In this case, the situation is complicated by the occurrence of an irregular singularity. The latter also requires a generalization of the presented isomonodromy method. In Section \ref{minv}, we give a brief introduction to isomonodromic deformations, where we find the exact form of the corresponding PVI or PV equation for a general Fuchsian equation with four or three regular singularities. In addition, we consider the transformation between Painlev\'e VI and Painelv\'e V by a coalescence of two of the singularities. In Section \ref{apps} we combine the results of the previous two sections to derive the exact forms of the corresponding Painlev\'e VI/V to the original Fuchsian differential equations of motion in Y$^{p,q}$ and its limits. Our findings are summarized in Section \ref{conc}.
%%%%%%%%%%%%%%%%%%%

\section{Sasaki-Einstein spaces $Y^{p,q}$} \label{ypqSect}

The primary focus of this work are the equations of motion for a scalar particle on the Y$^{p,q}$ geometry and other related backgrounds. The equations for Y$^{p,q}$ have been shown to reduce to a general Heun equation in \cite{Kihara:2005nt}, which is presented in detail in Section \ref{heunDer}. Here, we further consider the $q \rightarrow p$ transition resulting in Y$^{p, p}$ geometry. This limit is of interest since it effectively induces a coalescence of two of the singularities of the original Heun equation. From the CFT point of view, this corresponds to a phase transition of the quiver theory to a simpler one. In the subsequent subsections we study the Y$^{1,0}$ and Y$^{\infty,q}$ cases, showing that a similar coalescence also occurs.

\subsection{The $Y^{p,q}$ metric}
The line element of the five dimensional $Y^{p,q}$ space, parameterized by two positive coprime integers $p,q$ $(p>q)$, is written by~\cite{Gauntlett:2004yd,Martelli:2004wu}:
\begin{align}
	ds^2_{Y^{p,q}}=\frac{1-y}{6} d\Omega^2 + \frac{dy^2}{w(y)q(y)}  +
	\frac{q(y)}{9} \big( d\psi - \cos\theta d\phi \big)^2  + w(y) \Big[ d \alpha + f(y) \big(d\psi -\cos\theta d\phi \big)  \Big]^2,
	\label{eqn:Ypqmetric}
\end{align}
%
\begin{comment}
or in matrix form:
%
\begin{align}
	& G_{\mathbb{Y}^{p,q}}(y,\theta,\phi,\psi,\alpha) = \nonumber \\
	& \begin{pmatrix}
		\frac{1}{w\,q} & 0 & 0 & 0 & 0 \\
		0 & \frac{1-y}{6} & 0 & 0 & 0 \\
		0 & 0 &  \frac{1-y}{9}\sin^2+ \left(\frac{q}{9}+w\,f^2\right)\cos^2\theta & -\left(\frac{q}{9}-w\,f^2\right)\cos\theta & -w\,f\cos\theta \\
		0 & 0 & -\left(\frac{q}{9}-w\,f^2\right)\cos\theta & 
		\frac{q}{9}+ w\,f^2 & w\,f  \\
		0 & 0 & -w\,f\cos\theta  & w\,f & w
	\end{pmatrix}.
\end{align}
%
\end{comment}
%
where we have introduced the following definitions:
\begin{align} \label{wfDefs}
	&w(y) = \frac{2(b - y^2)}{1-y} ~,~~~
	q(y) =\frac{b-3y^2 +2y^3}{b-y^2}~,~~~
	f(y) = \frac{b - 2y + y^2}{6(b - y^2)}~,
 \\
	&b =  \frac{1}{2} - \frac{p^2 -3 q^2}{4p^3} \sqrt{4p^2 - 3 q^2}~, \quad d\Omega^2= d \theta^2 + \sin^2\!\theta\, d\phi^2.
\end{align}
The coordinates $\{y,\theta,\phi,\psi,\alpha\}$ span the following ranges:
\begin{eqnarray}
	y_1 \leq y \leq y_2~,~~~ 0 \leq \theta \leq \pi ~,~~~
	0 \leq \phi \leq 2 \pi~,~~~  0 \leq \psi \leq 2 \pi~,~~
	0 \leq \alpha \leq 2 \pi l ~. 
\end{eqnarray}
The period of $\alpha$ is $2 \pi l$ with
\begin{eqnarray}
	l = \frac{q}{3 q^2  - 2 p^2 + p \sqrt{ 4 p^2 - 3 q^2 }}.
\end{eqnarray}
The parameters $y_1$ and $y_2$ are the the two smallest roots of the cubic equation $b-3y^2 +2y^3=0$, i.e.
\begin{eqnarray}
	y_{1,2} = \frac{1}{4p} \left( 2 p \mp 3q - \sqrt{ 4 p^2 - 3 q^2 }
	\right),
\end{eqnarray}
with the remaining root given by:
\begin{eqnarray}
	y_3= \frac{3}{2} -(y_1 + y_2)
	= \frac{1}{2} + \frac{\sqrt{4p^2 -3q^2}}{2p} .
\end{eqnarray}

It is important to note that in the cases of a double root, e.g. $y_1 = y_2$, the range of the $y$ coordinate changes to $y_2 \leq y \leq y_3$.

\subsection{Scalar Laplacian on $Y^{p,q}$} \label{subScalar}

We consider the Schr\"odinger equation for a scalar field $\Phi(\mathbf{X})$ propagating on the $Y^{p,q}$ background\footnote{The operator $\Box$ is a Laplacian, not a d'Alembertian -- there is no time direction in $Y^{p,q}$. The time coordinate is explicitly accounted for by the energy $E$.},
\begin{equation}\label{eqSchr}
    \Box \Phi = -E\Phi,
\end{equation}
where the scalar Laplacian is given by
\begin{equation*}
    \Box\Phi=\frac{1}{\sqrt{|G|}} \,\partial_a \Big(\! \sqrt{|G|}\, G^{ab}\, \partial_b \Phi \Big), \quad a,b=1,2,3,4,5.
\end{equation*}
Here $G^{ab}$ are the components of the inverse Y$^{p,q}$ metric \eqref{eqn:Ypqmetric}, $\partial_a=\frac{\partial }{\partial X^a}$ and  $X^a = \{y,\theta,\phi,\psi,\alpha\}$. The resulting equation (\ref{eqSchr}) is the time-independent Schr\"odinger equation with energy $E$.

In order to separate the variables one has to look at the isometries of the background. The symmetries, as discussed in \cite{Martelli:2004wu}, are $SU(2)\times SU(2)\times U(1)$ and then, the Schr\"odinger equation should contain the corresponding Casimir operator $\hat{K}$ of the $SU(2)\times U(1)$ isometry (see Appendix \ref{appB}):
\begin{equation}
    \hat K=\frac{1}{\sin \theta}\frac{\partial}{\partial \theta}\bigg(\sin\theta\frac{\partial}{\partial \theta}\bigg)+\frac{1}{\sin^2\theta} \bigg(\frac{\partial}{\partial\phi}+\cos\theta \frac{\partial }{\partial\psi}\bigg)^2+\bigg(\frac{\partial}{\partial \psi}\bigg)^2,
\end{equation}
and also the Reeb Killing vector\footnote{ The operator $\hat{Q}_R$ represents the Reeb Killing vector field which corresponds to the ${\mathcal R}$-symmetry of the dual gauge theory
\cite{Martelli:2004wu}. In \cite{Martelli:2004wu} the Reeb vector is defined with different normalization, $\hat{Q}_R=3\partial_\psi -\partial_\alpha/2$.}, 
\begin{equation}
	\hat{Q}_R = 2 \frac{\partial}{\partial \psi} - \frac{1}{3} \frac{\partial}{\partial \alpha},
\end{equation}
corresponding to the ${\mathcal R}$-charge. Therefore, the Laplacian takes the form:
\begin{multline}
\square  = 
	\frac{1}{1 - y} \frac{\partial}{\partial y} \bigg(\!(1-y) wq \frac{\partial}{\partial y}\!\bigg) 
	\!+\! \left( \frac{3}{2} \hat{Q}_R \right)^2 
	\!+ \frac{1}{wq} \!\left(\frac{\partial}{\partial \alpha} + 3y \hat{Q}_R \!\right)^2  
	+ \frac{6}{1-y}\!\left[ \hat{K} - \left(\frac{\partial}{\partial \psi}  \right)^2 \right],
\end{multline}
where the functions $w(y)$ and $f(y)$ have been defined in \eqref{wfDefs}.

\subsection{Separation of variables and Heun equation along $y$} \label{heunDer}

Due to the isometry of the background, it is natural to assume the following ansatz for the wave function \cite{Kihara:2005nt}:
\begin{eqnarray}\label{eqWaveFans}
\Phi(y,\theta,\phi,\psi,\alpha) = \exp \bigg[ i\! \left( N_{\phi} \phi +
N_{\psi} \psi + \frac{N_{\alpha}}{l} \alpha \right)\!\bigg]
R(y)\, \Theta(\theta),
\end{eqnarray}
where  $N_{\phi} , N_{\psi} , N_{\alpha} \in {\mathbb Z}$.
Hence, along the $\theta$ direction Eq. (\ref{eqSchr}) reduces to
\begin{eqnarray}
\hat K\Theta(\theta) = - J(J+1)\Theta(\theta),
\label{ode-theta}
\end{eqnarray}
with $J$ corresponding to the $SU(2)$-spin quantum number. Similarly, along the $y$ direction, one finds the following second order ordinary differential equation:
\begin{eqnarray}
	&&\frac{1}{1-y}  \frac{d}{dy}\left[ (1-y) w(y)q(y) \frac{d}{dy}
	R(y)\right]
	- \left[\left(\frac{3}{2} Q_R \right)^2
	+
	\right. 
	\nonumber \\&&
	\frac{1}{w(y)q(y)} \left( \frac{N_{\alpha}}{l} + 3 y Q_R  \right)^2  
	\left. 
	+
	\frac{6}{1-y}\left( J(J+1) - N_{\psi}^2  \right) - E\right] R(y) = 0,
	\label{ode-y}
\end{eqnarray}
where $Q_R = 2 N_{\psi}-\frac{1}{3l}N_{\alpha}$ is the ${\mathcal R}$-charge  quantum number. Now it is easy to see that the second equation (\ref{ode-y}) is of Fuchsian-type with four regular singularities at
$y=y_1, y_2, y_3$ and $\infty$, i.e. it corresponds to the general Heun equation:
\begin{equation}
R''(y) + \left(\sum_{i=1}^3  \frac{1}{y-y_i} \right)\!  R'(y) +
v(y) R(y) =0.
\label{eqn-heun-y}
\end{equation}
In the expression above we have defined the following notations:
\begin{align}
	&v(y) = \frac{1}{H(y)}\left[ \mu - \frac{y}{4} E - \sum_{i=1}^3
	\frac{\alpha_i^2 H'(y_i)}{y -y_i}\right],~~~~~
	H(y) = \prod_{i=1}^3(y-y_i),
 \\
	&\mu = \frac{E}{4} - \frac{3}{2} J(J+1) + \frac{3}{32} \left( \frac{2}{3}
	\frac{N_{\alpha} }{l} - Q_R\right)^2,
\end{align}
and 
\begin{align}\label{eqAlpha1}
    \alpha_1&=\pm \frac{1}{4}\left[ N_\alpha\left( p+q-\frac{1}{3l} \right) -Q_R \right], \\
    \alpha_2&=\pm \frac{1}{4}\left[ N_\alpha\left( p-q+\frac{1}{3l} \right) +Q_R \right],\\\label{eqAlpha3}
    \alpha_3&=\pm \frac{1}{4}\left[ N_\alpha\left( \frac{-2p^2+q^2+p\sqrt{4p^2-3q^2}}{q}-\frac{1}{3l} \right) -Q_R \right].
\end{align}
It is convenient to write the energy in the form:
\begin{equation}\label{eqenergyKappa}
    E=4\kappa(\kappa+2),
\end{equation}
where $\kappa$ and $\kappa - 2$ are the characteristic exponents at $y = \infty$. Let us transform the singularities from $\{y_1, y_2, y_3, \infty\}$ to $\{0, 1, \tau=\frac{y_1-y_3}{y_1-y_2},\infty\}$ using the transformation
\begin{equation}
    z=\frac{y-y_1}{y_2-y_1},
\end{equation}
followed by the rescaling:
\begin{equation}
    R(z)=z^{|\alpha_1|} (1-z)^{|\alpha_2|} (\tau-x)^{|\alpha_3|} h(z).
\end{equation}
This yields the canonical form of the general Heun equation (\ref{eqStandardHeun}):
\begin{equation}\label{eqGHEYpq}
h''(z)+\left(\frac{\gamma }{z}+\frac{\delta }{z-1}+\frac{\epsilon }{z-\tau}\right)h'(z)+\frac{\alpha  \beta  z-k}{z (z-1) (z-\tau)} h(z)=0,
\end{equation}
with explicit parameters given by
\begin{align}
    &\alpha=-\kappa+\sum_{i=1}^3 |\alpha_i|, \quad \beta=2+\kappa+\sum_{i=1}^3 |\alpha_i|,  \\
    &\gamma=1+2|\alpha_1|, \quad \delta=1+2|\alpha_2|, \quad \epsilon=1+2|\alpha_3|.
\end{align}
One also has the expression for the accessory parameter,
\begin{equation}
    k=\big( |\alpha_1|+|\alpha_3| \big) \big( |\alpha_1|+|\alpha_3| +1\big) -|\alpha_2|^2 + \tau\left[ \big( |\alpha_1|+|\alpha_2| \big) \big( |\alpha_1|+|\alpha_2| +1\big) -|\alpha_3|^2 \right] - \tilde\mu,
\end{equation} 
where
\begin{align}
    &\tilde\mu=-\frac{\mu-y_1\kappa(\kappa+2)}{y_1-y_2}=\frac{p}{q} \left[ \frac{2}{3}(1-y_1)\kappa(\kappa+2) -J(J+1) +\frac{1}{16}\left( \frac{2N_\alpha}{3l}-Q_R \right)^2 \right].
\end{align}
Finally, the singularity parameter $\tau$ yields
\begin{equation}
    \tau=\frac{1}{2}+\frac{\sqrt{4p^2-3q^2}}{2q}.
\end{equation}

Consequently, we can obtain the characteristic exponents of the Heun equation (\ref{eqGHEYpq}). Following Eq. (\ref{eqHeuntoHeun}) one can write:

\begin{align}\label{eqSomeThetaEqs1}
&t=\tau=\frac{1}{2}+\frac{\sqrt{4p^2-3q^2}}{2q},
\\\label{eqSomeThetaEqs12}
&\theta _0=1-\gamma =-2|\alpha_1|,
\quad\theta _1=1-\delta=-2|\alpha_2|,
 \\\label{eqSomeThetaEqs13}
 &\theta _t=1-\epsilon=-2|\alpha_3|,
\quad \theta_\infty=\alpha-\beta=-2 (\kappa+1),
 \\\label{eqSomeThetaEqs2}
 &\kappa _1 =\alpha=-\kappa+\sum_{i=1}^3 |\alpha_i|,
\quad\kappa _2=\beta=2+\kappa+\sum_{i=1}^3 |\alpha_i|,
\end{align}
together with the following form of the Fuchs relation 
\begin{equation}\label{eqFuchsAgain}
\theta_0+\theta_1+\theta_t+\kappa_1+\kappa_2=2.
\end{equation}

\subsection{The limit $q\to p$ and $Y^{p,p}$ space} \label{yppHG}

In this subsection we consider the limit $q\to p$, which is explicitly forbidden by the original geometry constraint $p>q$. Therefore, the metric for Y$^{p,p}$ cannot be obtained by a direct application of this limit. It is also known that Y$^{p,p}$ is not a Sasaki-Einstein space. For instance, the Y$^{1,1}$ space possesses $A_{1}$ singularity, which is not present in Y$^{p,q}$ for $p > q$. Thus, it would be interesting to consider the formal limit $q\to p$ in the Heun equation \eqref{eqn-heun-y} found for $Y^{p,q}$. 

Before we consider the limit let us take a closer look at some special cases. As we already noted in the Introduction $Y^{1,1} , Y^{1,0}$ and $Y^{2,2}$ are
in a way exceptional as the only members of $Y^{p,q}$ family admitting massive supersymmetric deformations \cite{Feng:2000mi}.
This can be shown by using suspended pinch point (SPP) theory \cite{Aspinwall:1993nu} and holographic correspondence. Starting with the superpotential
\begin{equation}
W_{SPP}=X_{12}X_{23}X_{32}X_{21} - X_{23}X_{31}X_{13}X_{32} + X_{13}X_{31}X_{11} - X_{12}X_{21}X_{11},
\end{equation}
where $X_{ij}$ are bfundamentals. Higgsing $X_{12}$, i.e. $\langle X_{12}\rangle\neq 0$, makes $X_{12}$ and $X_{11}$ massive and, after integrating them out, gives the conifold superpotential consisting of two quartics. When $\langle X_{23}\rangle$, the superpotential becomes purely cubic resulting in $\mathbb{C}^3$ orbifold and enhanced supersymmetry to $\mathcal{N}=2$. This is illustrated in the toric diagram below.
\begin{figure}[H] 
\begin{center}
	\includegraphics[width=0.7\textwidth, trim={0.5cm 1.9cm 9cm 14cm},clip]{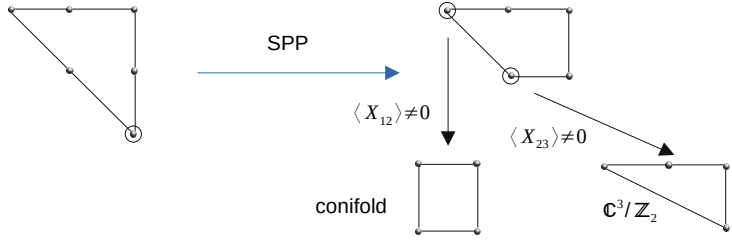}
	\end{center}
	\caption{SPP and obtaining toric diagrams of $T^{1,1}$ and $\mathbb{C}^3/\mathbb{Z}_2$ by Higgsing.}
\end{figure} \label{figToric}

The toric Calabi-Yau orbifold corresponding to the limit $q \to p$ in general is $\mathcal{X} := \mathbb{C}^3 /\mathbb{Z}_a$, where the group $\mathbb{Z}_a$ acts on $(z_1 , z_2 , z_3 ) \in \mathbb{C}^3$ as
\begin{equation}
(z_1 , z_2 , z_3 ) \to (\epsilon^1 z_1 , \epsilon^s z_2 , \epsilon^{-s-1} z_3 ),
\label{orb-1}
\end{equation}
with $\epsilon = e^{2\pi i/a}$, and toric weights $w := (w_1 , w_2 , w_3 ) = (1, s, -s -1),\, s \in \mathbb{Z}$, satisfy in general the condition $\sum_iw_i=0\,\operatorname{mod}a$. In particular, for $a=2$   $\mathbb{Z}_2$ acts on coordinates of $\mathbb{C}^3$ as
\begin{equation}
(z_1 , z_2 , z_3 ) \to (-z_1 , -z_2 , z_3 ), \quad \text{i.e.,\, $s = 1$ in \eqref{orb-1}}. 
\end{equation}
Specifically, the toric weights $(w_1,w_2,w_2)=(1,1,-2)$ and thus, for $Y^{1,1}$ the Calabi-Yau cone is $\mathbb{C}\times\mathbb{C}^2/\mathbb{Z}^2$ with an accidental $\mathcal{N}=2$ supersymmetry. 

The case of $p\neq 1$ and $q=0$, one can see that these are $\mathbb{Z}_p$ quotients of the cases above, namely $(\mathbb{C}\times\mathbb{C}^2/\mathbb{Z}_2)/\mathbb{Z}_p$ and $(\operatorname{conifold})/\mathbb{Z}_p$. It is interesting to note that the formal limit $p=q$ and $q=0$ is consistent with the general inequality for $Y^{p,q}$
\begin{equation}
\operatorname{vol}(T^{1,1}/\mathbb{Z}_p) > \operatorname{vol}(Y^{p,q}) > \operatorname{vol}(S^5/\mathbb{Z}_2\times \mathbb{Z}_p),
\end{equation}
although the metric for $Y^{p,q}$ is not anymore relevant for taking this limit. Due to the fact that when $q=p$, the parameter $b=1$, the base becomes weighted projective space and does permit the construction of a Sasaki-Einstein metric on such a space. The further analysis should be taken using toric description and gauge linear sigma model (GLSM) approach for instance.

Let us now take the formal limit $q\to p$ in the general Heun equation (\ref{eqGHEYpq}). In this case, there is a coalescence of singularities due to $t\to 1$ and also $\epsilon=\delta=1+|N_\psi|$. The result is the following hypergeometric equation (HypG):
\begin{align}\label{eqHGEabc}
  \nonumber & w''(z)+\frac{ \left( 2 J+3-| N_{\psi }-p N_{\alpha }|\right) z+| N_{\psi }-p N_{\alpha }| -1}{z (z-1) } w'(z) 
    \\
    &-\frac{ 4 \kappa  (\kappa +2)-4 J (J+2)+2 | N_{\psi }-p N_{\alpha }|  \big( 2 J+2-| N_{\psi }-p N_{\alpha }|\big)+\left(N_{\psi }-p N_{\alpha }\right)^2}{4 z (z-1) }w(z)=0,
\end{align}
where additional rescaling of the function has also been performed.
\begin{comment}
Consequently, assuming $N_{\psi }-p N_{\alpha }<0$ Eq. (\ref{eqHGEab}) with $r_{+}(z)$ and Eq. (\ref{eqHGEabc}) are the same for $L=-1-J$. Obviously, assuming $N_{\psi }-p N_{\alpha }>0$, Eq. (\ref{eqHGEab}) with $r_{-}(z)$ and Eq. (\ref{eqHGEabc}) are the same for $L=-1-J$. This is what we wanted to show.

However, there is another hypergeometric equation coming from GHeun if one assumes $2 \kappa +1-2 J>0$. It can be obtained if we replace $J\to -J-1$ in Eq. (\ref{eqHGEabc}) {\color{red}{check this}}. In this case it coincides with Eq. (\ref{eqHGEab}) for $L=J$.
\end{comment}
We can compare the above equation to the standard form of the hypergeometric one from (\ref{EqGaus1}) and extract the coefficients $a, b$ and $c$:
\begin{align}
    &a= \pm(1+\kappa)-J-\frac{1}{2} | N_{\psi }-p N_{\alpha }| ,
    \\
    &b=  \mp(1+\kappa)-J-\frac{1}{2} | N_{\psi }-p N_{\alpha }|,
    \\
    &c= 1-| N_{\psi }-p N_{\alpha }|.
\end{align}
This leads to the following characteristic exponents:
\begin{align} \label{ppHypTh}
   & \theta_0^{(hyp)}=\bar \theta_0=1-c=| N_{\psi }-p N_{\alpha }|,
   \\
   &\theta_1^{(hyp)}=\bar \theta_t=c-a-b=1+2 J,
   \\
   &\theta_\infty^{(hyp)}=\bar \theta_\infty=a-b=\pm 2(1+\kappa).
\end{align}

However, these are not direct limits of the original characteristic exponents $\theta_{0,1,t,\infty}$ from (\ref{eqSomeThetaEqs12})-(\ref{eqSomeThetaEqs13}), which at $q\to p$ yields
\begin{equation}\label{eqThetasQtoP}
    \theta _0^{(q\to p)}= -| N_{\psi }-p N_{\alpha }|,\quad \theta _1^{(q\to p)}=\theta_t^{(q\to p)} =-| N_{\psi }|,\quad \theta _{\infty }^{(q\to p)}= -2 (\kappa +1).
\end{equation}
The reason for this is taking the limit $Y^{p,q} \rightarrow Y^{p,p}$ is non-trivial as explained earlier in the section. It is important to note, that these numbers still inherit the quantum integer nature of the original space. 

This limit of the geometry has an interesting effect on the characteristic exponents of the resulting hypergeometric equation. If we take a closer look at $\theta_0^{(hyp)}$ and $\theta_1^{(hyp)}$ we see that there are cases where both of the are non-zero integers\footnote{The field over which the differential equation is defined is $\mathbb{C}_3 / Z_2$, based on the toric diagrams discussed earlier in this section}. Specifically, $\theta_0^{(hyp)}$ is always 1 over this field, so this singular point is always resonant\footnote{In this context a resonant singular point is defined as having two characteristic exponents whose difference is a non-zero integer}. The case for $\theta_1^{(hyp)}$ is a little more complicated and it can be separated into two main categories. If $p$ is even the resonance of the point depends on $N_\psi$, whereas if $p$ is odd the resonance will depend on whether the sum $N_\psi - N_\alpha$ is odd. The cases when each of the singularities is resonant should be considered separately from the cases of non-resonant points. This is due to the fact that the Frobenius manifold formed by the monodromy data has different topology, which implies that the derivations in sections \ref{sectPVI} and \ref{sectPV} are not generally applicable. For the appropriate treatment in the context of isomonodromic deformations for resonances, please consult \cite{bolibruch1998isom,bertola2005isom, cotti2019isom}. The corresponding analysis of the associated Painlev\'e V equations by sectors is beyond the scope of this work. 

\subsection{Y$^{1,0}$ and the T$^{1,1}/\mathbb{Z}_2$ limit}
As shown in \cite{Gauntlett:2004yd}, in certain cases Y$^{p,q}$ space can be reduced to T$^{1,1}$ or other related to T$^{1,1}$ geometries. In this section, we will consider the limit Y$^{1,0}$ in which the geometry reduces to T$^{1,1} / \mathbb{Z}_2$. Scalar particles on these geometries have been studied in some works, e.g. \cite{us2020pulsT11, arnaudov2011pulsating}, however the method of isomonodromic deformations has not been applied to the resulting equations. For completeness, we also derive the relevant Fuchsian equations.
Let us consider the standard T$^{1, 1}$ metric of the form:
\begin{equation}
dS_{T^{1,1}}^2 = \frac{b}{4} \Big( d\theta_1^2 + \sin^2\!\theta_1\, d\phi_1^2 + d\theta_2^2 + \sin^2\!\theta_2\, d\phi_2^2 + b \big( d\psi - \cos\theta_1\, d\phi_1 - \cos\theta_2\, d\phi_2 \big)^2 \Big),
\end{equation}
where $b = \frac{2}{3}$ and $\theta_1, \theta_2, \phi_1, \phi_2, \psi$ are the coordinates of the space. Equivalently to the analysis for Y$^{p,q}$ in Section \ref{subScalar}, we consider the Schr\"odinger equation on $T^{1,1}$. Due to the isometries of the space, we can suppose an ansatz of the form
\begin{equation}
    \Phi = e^{i(N_1 \phi_1 + N_2 \phi_2 + N_\psi \psi)}\, T_1(\theta_1) T_2(\theta_2),
\end{equation}
where $N_1, N_2, N_\psi$ are integers and $T_1, T_2$ are arbitrary functions. After separation of variables we obtain two equations for $\theta_1$ and $\theta_2$ in the form:
\begin{equation}
T_i''(\theta_i) + \cot\theta_i\, T_i'(\theta_i) + \bigg( {E_i} - \frac{1}{\sin^2\!\theta_i} \big(N_\psi \cos\theta_i +N_i\big)^2\bigg) T_i(\theta_i) = 0, \quad i \in \{ 1,2\},
\end{equation}
where the separation constants $E_i$ must satisfy the additional condition:
\begin{equation}
    E = \frac{4}{b^2} \big( b E_1 + b E_2 + N_\psi^2 \big). 
\end{equation}
Upon substituting new variables $x_i = \cos
\theta_i$ and rescaling the variable to a new one $z_i = \frac{1}{2}(x_i+1)$, we obtain two standard hypergeometric equations of the form:
\begin{equation} \label{hgT11}
    z_i(z_i-1)T_i''(z_i) + \big( N_i - N_\psi - 1 + 2z_i(1+N_\psi) \big) T_i'(z_i) + \big( N_\psi - E_i \big) T_i(z_i) = 0.
\end{equation}
These equations have the following characteristic exponents:
\begin{equation} \label{thY10}
\theta^{(i)}_0 = N_i - N_\psi - 1, \quad \theta^{(i)}_1 = N_i -2 - 3 N_\psi, \quad \theta^{(i)}_\infty = -\sqrt{1 + 4 N_\psi + \frac{E_i}{b^2}},
\end{equation}
where $\theta^{(i)}_{z_i}$ signifies the characteristic exponent of the $i$-th equation with respect to the expansion at singular point $z_i$. It is interesting to note that unlike the result obtained for the general Y$^{p,q}$ metric, here the equation reduces to two hypergeometric equations whose coefficients have an algebraic relation between each other. This can be thought as a result of the fact that in the $ p = 1, q = 0$ limit the symmetry group of the original space changes from $SU(2) \times SU(2) \times U(1)$ for Y$^{p,q}$ to $SU(3) \times SU(2)$ for $T^{1,1} / \mathbb{Z}_2$. Once again we need to be aware of situations in which $\theta^{(i)}_0$ and $\theta^{(i)}_1$ are resonant. Since our parameters are defined over $\mathbb{Z}_2$ the cases of odd characteristic exponents are resonant. This is similar to the phenomenon discussed in the previous section, i.e. based on the choice of quantum numbers, it is possible for singular points to alternate between resonant and non-resonant. The effect of the resonant  contributions to the full spectrum of generated Painlev\'e V equations is beyond the scope of this work and will be considered in a separate paper.  \\

\subsection{The Y$^{\infty, q}$ limit}

Here we show that Y$^{\infty, q}$ is a confluent\footnote{By confluent, we mean that such limits induce a confluence of some of the singularities of the original geometry, thus leading to changes in the topology of the Frobenius manifolds of the corresponding equations of motion. This will result in a confluent Heun equation} limit of the original Y$^{p,q}$ geometry. Let us consider the original metric \eqref{eqn:Ypqmetric}. Taking the limit $p\to\infty$ in $b$ and $l$, we get:
\begin{equation} \label{infLimits}
\lim_{p \to \infty} b = 0, \quad \lim_{p \to \infty} l = \frac{4}{9 q}.
\end{equation}
For the other functions in the metric we find:
\begin{equation} \label{infFuncs}
\lim_{p \to \infty} w(y) = \frac{2y^2}{y-1},\quad \lim_{p \to \infty} q(y) = 3 - 2y, \quad \lim_{p \to \infty} f(y) = \frac{2-y}{6y}.
\end{equation}
Now the separation ansatz for the wave function in the Schr\"odinger equation is assumed to be:
\begin{align}
\Phi_{p \to \infty} = \exp{\bigg[i \bigg( N_\phi \phi + N_\psi \psi + \frac{9 q N_\alpha}{4} \alpha\bigg) \bigg] T(\theta) Y(y) }.
\end{align}
For the $\theta$ component of the equation we obtain the standard Jacobi equation with eigenvalues $J(J+1)$, $J \in \mathbb{Z}$. Rescaling $y \to \frac{2}{3} y$, the $y$ equation becomes:
\begin{align} \label{infEqY1}
&Y''(y) + \bigg(\frac{1}{y-1} + \frac{2}{y}\bigg) Y'(y) + \frac{p_4(y)}{768(y-1)^2 y^4} Y(y) = 0, \\ \nonumber
&p_4 (y) = 192 E y^4 + \bigg( 320 E - 128 J(J+1) + 3( 8 N_\psi - 9 N_\alpha q)^2 \bigg) y^3 \\ 
&+2 \bigg(64 J(J+1) - 3( 8 N_\psi - 9 N_\alpha q)^2 + 320 E \bigg) z^2 + 96 N_\alpha q (3 N_\alpha q - 4 N_\psi)-48 N_\alpha^2 q^2.
\end{align}
To write this equation in a standard form, we change variables to $y = \frac{1}{z}$ and then rescale. Hence, we obtain the final form of our confluent Heun equation:
\begin{align}
    &h''(z) + \bigg( \frac{1 + \sqrt{4+E}}{z} + \frac{8 + |N_\alpha q - 8 N_\psi|}{8 (z-1)} + \frac{N_\alpha q}{2} \bigg) h'(z) + \frac{p_1(z)}{z(z-1)} h(z) = 0,\\
    \nonumber
    &p_1(z) = 1 + \frac{1}{6} J(J+1) - \frac{1}{4} (N_\psi^2 - N_\alpha q) + \frac{9 N_\alpha q} {16} \bigg(q - \frac{9 N_\alpha q}{16} \bigg) + \frac{E}{12} \\
    \nonumber
    &+\frac{4+E}{2} \bigg(1 - \frac{N_\alpha q}{4} + \frac{|N_\alpha q - 8 N_\psi|}{16} \big(1 + \sqrt{4+E}\,\big)\! \bigg) \\
    &+z \frac{N_\alpha q}{32} \big( 16 - 16 N_\psi + 8 N_\alpha + \sqrt{4+E} +  |N_\alpha q - 8 N_\psi| \big)
\end{align}
This confluent Heun equation has two regular singularities at $0, 1$ and an irregular singularity of rank 1 at $\infty$. This implies that the characteristic exponents of the equations are:
\begin{align} \label{thYinf}
    \theta_0 = -\sqrt{4+E}, \quad \theta_1 = -\frac{1}{8} |N_\alpha q - 8 N_\psi|,\\ \quad \theta_\infty = 1 - N_\psi + N_\alpha q + \frac{1}{16} \Big( \sqrt{4 + E} + |N_\alpha q - 8N_\psi| \Big). 
\end{align}

The presence of an irregular singularity at $\infty$ indicates that the isomonodromic methods derived below, will not be applicable to this case. This situation leads to a separation of the monodromy space into Stokes sector. Similarly to the case of resonant points, the latter phenomenon goes beyond the scope of this work and will be considered in a separate study.

\section{Isomonodromic Deformation of Fuchsian equations} \label{minv}

In this section we will give a brief outline of the method of isomonodromic deformation and apply it to ODEs with three and four regular singular points, which are equivalent to the ones derived in the previous section. For more detail, the reader is advised to consult \cite{iwasaki:1991, conte:2006, haraoka:2015}. The main motivation, apart from allowing us to find asymptotic solutions to the equations, it allows us to study and compare the geometries from the view point of their associated Painlev\'e equations \ref{appPainleve}. Furthermore, the coalescence cascade of the Painlev\'e equations allows us to compare geometries even if the two geometries do not have the same corresponding Painlev\'e equation.

\subsection{General setup} \label{isomTh}
Any second order Fuchsian system can be represented as a system of first order ODEs as:
\begin{equation} \label{dzA}
\partial_z \mathbf{y}(z) = A(z) \mathbf{y}(z),\quad A(z) = \sum_{i} \frac{A_i(z)}{z - t_i},
\end{equation}
where $\mathbf{y}$ is a $2d$ vector composed of linearly independent solutions and $t_i$ are the set of singular points of our differential equation. It is important to note that in general we can consider the set of singularities $0, 1, t_i, \infty$, where $t_i$'s can be of arbitrary number. We are free to do so as we consider a one-dimensional complex projective space $\mathbb{P}_1(\mathbb{C})$ and thus we can the M\"obius transform, that is an automorphism on this space, to fix any three singularities. It is convenient that the example we are considering has 4 singularities, as in general systems with 5 or more singular do not admit monodromy-preserving families of solutions. For a system to admit such solutions, $\mathbf{y}(z)$ needs to satisfy both \eqref{dzA} and the auxiliary equation:
\begin{equation} \label{auxA}
d_{t_i} \mathbf{y}(z) = \Omega \mathbf{y}(z),\quad \Omega = \Omega_i(z,t_i) dt_i,\quad  t_i = {0, 1, t}.
\end{equation}
Furthermore, if we require the system of equations \eqref{dzA} and \eqref{auxA} to be integrable by the Pfaffe condition. This means that $\Omega(z,t_i)$, $A(z,t_i)$ must satisfy the following equations:
\begin{equation} \label{pfaffe1}
d_{t_i}A = \partial_z\Omega - [A, \Omega],
\end{equation}
where
\begin{equation} \label{pfaffe2}
d_{t_i}\Omega = \Omega \wedge \Omega.
\end{equation}
If we assume that the no two characteristic exponents differ by as integer and the connection at infinity can be represented as:
\begin{equation}
    A_{\infty} = \begin{pmatrix}
            \kappa_1-1 & 0 \\
            0  & \kappa_2-1
            \end{pmatrix},
\end{equation}
then it can be shown that the components of $\Omega$ satisfy:
\begin{equation}
    \Omega_i = -\frac{A_i}{z - t_i}.
\end{equation}
Plugging this representation into the Pfaffian conditions of integrability \eqref{pfaffe1} and \eqref{pfaffe2}, we obtain the well known form of the Schlesinger equations:
    \begin{align}\label{schlesinger}
        \partial_{t_i} A_j = \frac{[A_i, A_j]}{t_i - t_j} \;\;\; i \neq j,\quad 
        \partial_{t_i} A_i = -\sum_{i \neq j} \frac{[A_i, A_j]}{t_i - t_j} \;\;\; i = j.
    \end{align}
Another important property can be noticed when writing \eqref{schlesinger} as a single expression:
\begin{equation} \label{longschl}
    \frac{A_i}{\partial t_j} = \bigg[A_i, (1 - \delta_{ij}) \frac{A_j}{t_i - t_j} + \delta_{ij} \sum_{j \neq i} \frac{A_i}{a_i - a_j} \bigg].
\end{equation}
From this it can be seen that the Schlesinger equations are invariant with respect to conjugation of $A$ matrices (i.e. $A \to gAg^{-1}$), which means that the submanifold of each moving $t_i$ is equipped with a symplectic form and thus possesses Hamiltonian structure. Alternatively, expression \eqref{longschl} can be immediately read off as the general expression of a symplectic form on a coadjoint orbit. Regardless, this means that on each submanifold there exists a pair of conjugate coordinates $\lambda_i, \mu_i$ and a generating Hamiltonian $K_i$ that govern the isomonodromic flows as:
    \begin{align} \label{poissonGEN}
        \frac{\partial \mu_i}{\partial t_j} = \{K_j, \mu_i \},\quad
        \frac{\partial \lambda_i}{\partial t_j} = \{K_j, \lambda_i \},
    \end{align}
where the $\{, \}$ is just the standard Poisson bracket defined as:
\begin{equation} \label{poisson}
    \{f, g\} = \frac{\partial f}{\partial \lambda_i} \frac{\partial g}{\partial\mu_i} - \frac{\partial f}{\partial \mu_i} \frac{\partial g}{\partial\lambda_i},
\end{equation}
It turns out that the canonical variables $\lambda_i$ are actually just the set of apparent singularities for our equation. Furthermore, it was shown in the 1980s works of the Kyoto school that the generating Hamiltonians are related to the $\tau$-function as \cite{iwasaki:1991}:
\begin{equation} \label{Htau}
    K_i = \frac{\partial \ln \tau}{\partial t_i} dt_i = \sum_{i \neq j} \frac{\tr A_i \tr A_j}{t_i - t_j} - \frac{\tr(A_i A_j)}{t_i - t_j},
\end{equation}
which can also be represented as:
\begin{equation} \label{resH}
    K_i = -\mathop{{\rm Res}}_{z \rightarrow t_i} Q(z),
\end{equation}
where $Q(z)$ is the coefficient of $y(z)$ in our original second order ODE. A similar expression holds for the other conjugate variable $\mu$, which can be represented as:
\begin{equation} \label{resmu}
    \mu_i = A_{11}(\lambda_i) = \mathop{{\rm Res}}_{z \rightarrow \lambda_i} Q(z),
\end{equation}
which can be obtained by the Sklyanin method for separation of variables and finding Darboux coordinates \cite{sklyanin:1995}. On the other hand, there exists a natural parametrization for each component of our connection matrices $A_i$. Since these matrices are effectively a member of a $gl(2)$ symplectic manifold their components can be represented as\footnote{This parametrization is not unique. However, we use it since for the Heun equation the matrices $A_i$ must satisfy the last two relations of \eqref{eqAparam}}:
\begin{equation}
    A_i = \begin{pmatrix}
        p_i + \theta_i & -p_i q_i \\
        \frac{1}{q_i}(p_i + \theta_i) & -p_i
           \end{pmatrix},
\end{equation}
where $p_i, q_i$ are not the same conjugate coordinates as $\lambda_i, \mu_i$ (cite the papers with the derivation with Lie-Poisson brackets), but they can sometimes be related. In the general case finding this relation is not possible, but in the case of four singular possible it always is. Furthermore, if we find said connection, we can then derive the equations that govern the flow of $\mu_t, \lambda_t$, we obtain the sixth Painlev\'e equation. This calculation is done in the following section. 

\subsection{The Schlesinger system for Heun equation and Painlev\'e VI} \label{sectPVI}

Let us write a homogeneous linear ODE of order 2 as a set of 2 coupled linear ODEs of first order:
\begin{equation}
    y_i^{\prime}(z)=\sum\limits_{j=1}^2 A_{ij}(z) y_j(z), 
\end{equation}
or explicitly
\begin{align}\label{eqY1Prime}
    &y_1^{\prime}=A_{11} y_1+A_{12} y_2,
    \\\label{eqY2prime}
    &y_2^{\prime}=A_{21} y_1+A_{22} y_2.
\end{align}
Let us take $\partial_z$ from the second equation:
\begin{equation}\label{eqY2Second}
    y_2^{\prime\prime}=A_{21}^{\prime} y_1+A_{21} y_1^{\prime}+A_{22}^{\prime} y_2+A_{22} y_2^{\prime}.
\end{equation}
We express $y_1$ from (\ref{eqY2prime}):
\begin{equation}
    y_1=\frac{y_2^{\prime}-A_{22} y_2}{A_{21}}.
\end{equation}
By substituting it into (\ref{eqY2Second}), together with $y_1^{\prime}$ from (\ref{eqY1Prime}), one finds
\begin{equation}\label{eqY2Seconda}
    y_2^{\prime\prime}(z)-\big({\rm{Tr}}A+\partial_z\ln A_{21}\big) y_2^{\prime}(z)+\big(\det A-A_{22}^{\prime}+A_{22} \partial_z\ln A_{21}\big) y_2(z)=0.
\end{equation}
Similar equation is valid also for $y_1(z)$:
\begin{equation}\label{eqY1Seconda}
    y_1^{\prime\prime}(z)-\big({\rm{Tr}}A+\partial_z\ln A_{12}\big) y_1^{\prime}(z)+\big(\det A-A_{11}^{\prime}+A_{11} \partial_z\ln A_{12}\big) y_1(z)=0,
\end{equation}
where
\begin{equation}
    {\rm{Tr}}A=A_{11}+A_{22},\quad \det A=A_{11} A_{22}-A_{12} A_{21},\quad \partial_z\ln A_{21}=\frac{A_{21}^{\prime}}{A_{21}},\quad \partial_z\ln A_{12}=\frac{A_{12}^{\prime}}{A_{12}}.
\end{equation}
Obviously Eq. (\ref{eqY2Seconda}) appears to have an additional new singularity at $A_{21}(z)=0$, thus $z=\lambda$, or similarly Eq. (\ref{eqY1Seconda}) -- at $A_{12}(z)=0$, thus again one may write $z=\lambda$. Let us consider only Eq. (\ref{eqY1Seconda}) and loose the subscript for $y_1$:
\begin{equation}\label{eqY1Secondb}
    y^{\prime\prime}(z)-\big({\rm{Tr}}A+\partial_z\ln A_{12}\big) y^{\prime}(z)+\big(\det A-A_{11}^{\prime}+A_{11} \partial_z\ln A_{12}\big) y(z)=0.
\end{equation}
It is convenient to write this equation such as
\begin{equation}\label{eqDeformedHeun}
     y^{\prime\prime}(z)+P(z) y^{\prime}(z)+ Q(z) y(z)=0.
\end{equation}
where
\begin{equation}\label{eqPQz}
    P(z)=-{\rm{Tr}}A-\partial_z\ln A_{12},\quad Q(z)=\det A-A_{11}^{\prime}+A_{11} \partial_z\ln A_{12}.
\end{equation}

Because $A_{12}(z)=0$ at $z=\lambda$, one may choose $A_{12}(z)$ such as
\begin{equation}\label{eqA12a}
    A_{12}(z)=\frac{\zeta (z-\lambda)}{z (z-1)(z-t)}, \quad \zeta=const.
\end{equation}

Let us consider the following form of the matrix $A(z)$:
\begin{equation}\label{eqAmatrix}
    A(z)=\left(
\begin{array}{cc}
 A_{11}(z) & A_{12}(z)  \\
 A_{21}(z) & A_{22}(z) \\
\end{array}
\right)=\sum\limits_{i=0}^t\frac{A_i}{z-z_i}=\frac{A_0}{z}+\frac{A_1}{z-1}+\frac{A_t}{z-t},
\end{equation}
where $z_i=(0,1,t)$ and the constant matrices $A_i$ can be parameterized such as:
\begin{equation}\label{eqAparam}
    A_i=\left(
\begin{array}{cc}
 p_i+\theta_i & -p_i q_i  \\
 \frac{p_i+\theta_i}{q_i} & -p_i \\
\end{array}
\right)=\left(
\begin{array}{cc}
 A_{i,11} & A_{i,12}  \\
 A_{i,21} & A_{i,22} \\
\end{array}
\right), \quad {\rm{Tr}} A_i=\theta_i,\quad \det A_i=0,
\end{equation}
where $p_i$ and $q_i$, $i=0,1,t$, are some parameters as functions of $(\theta_0,\theta_1,\theta_t,\lambda,\mu,\kappa_1,\kappa_2)$. This leads to the following components for the matrix $A$:
\begin{align}\label{eqA112bb}
   & A_{11}(z)=\frac{A_{0,11}}{z}+\frac{A_{1,11}}{z-1}+\frac{A_{t,11}}{z-t}=\frac{p_0+\theta_0}{z}+\frac{p_1+\theta_1}{z-1}+\frac{p_t+\theta_t}{z-t},
    \\\label{eqA12b}
   & A_{12}(z)=\frac{A_{0,12}}{z}+\frac{A_{1,12}}{z-1}+\frac{A_{t,12}}{z-t}=-\frac{p_0 q_0}{z}-\frac{p_1 q_1}{z-1}-\frac{p_t q_t}{z-t},
   \\
& A_{21}(z)=\frac{A_{0,21}}{z}+\frac{A_{1,21}}{z-1}+\frac{A_{t,21}}{z-t}=\frac{p_0+\theta_0}{q_0 z}+\frac{p_1+\theta_1}{q_1(z-1)}+\frac{p_t+\theta_t}{q_t (z-t)},
\\
& A_{22}(z)=\frac{A_{0,22}}{z}+\frac{A_{1,22}}{z-1}+\frac{A_{t,22}}{z-t}=-\frac{p_0}{ z}-\frac{p_1}{z-1}-\frac{p_t}{ z-t}.
\end{align}

We can choose the matrix at infinity to be in a diagonal form:
\begin{equation}\label{eqAinfty}
    A_\infty=-(A_0+A_1+A_t)=\left(
\begin{array}{cc}
 \kappa_1-1 & 0 \\
 0 & \kappa_2-1 \\
\end{array}
\right).
\end{equation}
Substituting parameterized form (\ref{eqAparam}) for the $A_i$ in the above equation, one has the following conditions:
\begin{align}\label{eqP0P1P2K1}
    &p_0+p_1+p_t+\theta_0+\theta_1+\theta_t=1-\kappa_1,
    \\\label{eqP0P1Ptq0q1qt}
    &p_0 q_0+p_1 q_1+p_t q_t=0,
    \\\label{eqP0P1Ptq0q1qtt1t2tt}
    &\frac{p_0+\theta_0}{q_0}+\frac{p_1+\theta_1}{q_1}+\frac{p_t+\theta_t}{q_t}=0,
    \\\label{eqP0P1P2K2}
    &p_0+p_1+p_t=\kappa_2-1.
\end{align}
Obviously, plugging the last equation into the first one we recover the Fuchsian condition (\ref{Fuch_rel}) for Heun's equation: 
\begin{equation}
    \theta_0+\theta_1+\theta_t+\kappa_1+\kappa_2=2.
\end{equation}
Therefore, only three of the equations above are independent. For example we can eliminate one of the $p_0$, $p_1$ or $p_t$ from either Eq. (\ref{eqP0P1P2K1}) or Eq. (\ref{eqP0P1P2K2}).

Now, let us find out the equations for $\zeta$ and $\lambda$ coming from $A_{12}$ (Eqs. (\ref{eqA12a}) and (\ref{eqA12b})):
\begin{equation}
    A_{12}=\frac{\zeta (z-\lambda)}{z (z-1)(z-t)}=-\frac{p_0 q_0}{z}-\frac{p_1 q_1}{z-1}-\frac{p_t q_t}{z-t}.
\end{equation}
After finding the common denominator of the right hand side and comparing the coefficients in front of $z^0,\,z^1$ and $z^2$, one finds the following two relations:
\begin{align}\label{eqkp0q0lt}
   &p_0 q_0 t - \zeta \lambda = 0,
   \\\label{eqkp0q0p1q1ptqtt}
   &\zeta-p_0 q_0 (t+1)-p_1 q_1 t-p_t q_t=0, \\
   &p_0 q_0+p_1 q_1+p_t q_t=0,
\end{align}
where the last equation coincides with (\ref{eqP0P1Ptq0q1qt}). With the chosen form of the matrix $A$ (\ref{eqAmatrix}) and the explicit parametrization of $A_i$ (\ref{eqAparam}), one can write $P(z)$ and $Q(z)$ from (\ref{eqPQz}) in the form \cite{amado:2020}:
\begin{equation} \label{PPVI}
    P(z)=\frac{1-\tr A_0}{z}+\frac{1-\tr A_1}{z-1}+\frac{1-\tr A_t}{z-t}-\frac{1}{z-\lambda},
\end{equation}
\begin{align}
\nonumber   Q(z)&=-\frac{t (t-1)}{z (z-1)(z-t)} \bigg(\frac{A_{0,11}+A_{t,11}}{t}+\frac{A_{1,11}+A_{t,11}}{t-1}+\frac{A_{t,11}}{\lambda-1}\bigg)+\frac{\lambda (\lambda-1) A_{11}(z)}{z (z-1)(z-\lambda)}
   \\
   &+\frac{\det A_0}{z^2}+\frac{\det A_1}{(z-1)^2}+\frac{\det A_t}{(z-t)^2}
   +\frac{\varrho}{z (z-1)}-\frac{t (t-1) H}{z (z-1)(z-t)}+\frac{A_{\infty,11}}{z (z-1)},
\end{align}
where $A_{11}(z)$ is defined in (\ref{eqA112bb}) and
\begin{align}
    \varrho=\det A_\infty-\det A_0-\det A_1-\det A_t=(\kappa_1-1) (\kappa_2-1),
\end{align}
\begin{equation}
    H=-\frac{1}{t} {\rm{Tr}} A_0 {\rm{Tr}} A_t-\frac{1}{t-1} {\rm{Tr}} A_1 {\rm{Tr}} A_t+\frac{1}{t} {\rm{Tr}}(A_0 A_t)+\frac{1}{t-1} {\rm{Tr}}(A_1 A_t).
\end{equation}

We still need to introduce $\mu$, which is the conjugate variable to $\lambda$ and also their Hamiltonian $K$, where we follow the notations by \cite{amado:2020}. By definition one has \cite{iwasaki:1991}:
\begin{align}\label{eqMu}
 \mu&=\mathop{\rm{Res}}_{z=\lambda} Q(z) =A_{11}(\lambda)=\frac{A_{0,11}}{\lambda}+\frac{A_{1,11}}{\lambda-1}+\frac{A_{t,11}}{\lambda-t}
  =\frac{\theta _0+p_0}{\lambda }+\frac{\theta _1+p_1}{\lambda -1}+\frac{p_t+\theta _t}{\lambda -t},
\end{align}
and the Hamiltonian $K$:
\begin{align}\label{eqHa}
K&=-\mathop{\rm{Res}}_{z=t} Q(z)=H+\frac{\lambda (\lambda-1)\mu}{t (t-1)}+\frac{(\lambda-t) A_{\infty,11}}{t (t-1)}.
\end{align}

The seven equations \eqref{eqP0P1Ptq0q1qt}, 
 \eqref{eqP0P1Ptq0q1qtt1t2tt}, 
 \eqref{eqP0P1P2K2}, \eqref{eqkp0q0lt}, \eqref{eqkp0q0p1q1ptqtt}, \eqref{eqMu} and \eqref{eqHa} are sufficient to find the parameters $(\zeta, p_0,p_1,p_t,q_1,q_t)$. Only $q_0$ is still arbitrary. This is going to cause no problem, because $q_0$ will drop out naturally from the final form of the Hamiltonian $K$. To make things as easy as possible we explain the procedure for solving these equations in details. First we solve Eq. (\ref{eqkp0q0lt}) with respect to $\zeta$:
 \begin{eqnarray}
     \zeta=\frac{p_0 q_0 t}{\lambda }.
 \end{eqnarray}
Next, we insert $\zeta$ in Eq. (\ref{eqkp0q0p1q1ptqtt}) and solve it  with respect to $q_t$:
\begin{equation}
    q_t= -\frac{p_0 q_0 (\lambda +(\lambda -1) t)+\lambda  p_1 q_1 t}{\lambda  p_t}.
\end{equation}
Now, we substitute $q_t$ in Eq. (\ref{eqP0P1Ptq0q1qt}) and solve it with respect to $q_1$:
\begin{equation}
    q_1= -\frac{(\lambda -1) p_0 q_0 t}{\lambda  p_1 (t-1)}.
\end{equation}
In the next step we solve Eq. (\ref{eqMu}) with respect to $p_t$:
\begin{equation}
    p_t= (\lambda-t ) \left( \mu-\frac{\theta _0+p_0}{\lambda }-\frac{\theta _1+p_1}{\lambda -1}-\frac{\theta _t}{\lambda -t}\right).
\end{equation}
Hence we can substitute it in Eq. (\ref{eqP0P1P2K2}) and solve it with respect to $p_1$:
\begin{align}
    p_1= \frac{(1-\lambda ) p_0 t+(\lambda -1) \lambda  \left(\kappa_2-\lambda  \mu +\theta _t+\mu  t-1\right)+\theta _0 (\lambda -1) (\lambda -t)+\theta _1 \lambda  (\lambda -t)}{\lambda  (t-1)}.
\end{align}
Using the expressions of $q_t, q_1, p_t$ and $p_1$ in this order, together with $\theta _0+\theta _1+\theta _t\to -\kappa _1-\kappa _2+2$, we can solve Eq. (\ref{eqP0P1Ptq0q1qtt1t2tt}) with respect to $p_0$:
\begin{align}
  \nonumber  p_0&= \frac{1}{t\big(\kappa _1-\kappa _2\big) }\bigg[\lambda  \left(\kappa _1+\lambda  \mu -1\right) \big(\theta _0 (1-\lambda )-\theta _1 \lambda +(\lambda -1) \left(\lambda  \mu-\kappa _2 -\theta _t+1\right)\big)
    \\ \nonumber
    &-t \bigg\{\theta _0^2 (\lambda -1)+\theta _0 \big(\theta _1 (2 \lambda -1)+(\lambda -1) \left(2 \kappa _2-2 \lambda  \mu +\theta _t-2\right)\big)
    \\ \nonumber
    &+\lambda  \bigg(\theta _1^2+\kappa _2 \left(\kappa _2- (\lambda -1) 2 \mu -2\right)+\mu(\lambda -1)   (\lambda  \mu +2)+\theta _1 \left(2 \kappa _2-(2 \lambda -1) \mu +\theta _t-2\right)
    \\
    &+\theta _t \left(\kappa _2-(\lambda  -1)\mu -1\right)+1\bigg)\bigg\}\bigg].
\end{align}
The final step is to substitute $q_t, q_1, p_t, p_1$ and $p_0$ in this order, together with $\theta _0+\theta _1+\theta _t\to -\kappa _1-\kappa _2+2$, in Eq. (\ref{eqHa}) to find the Hamiltonian $K$:
\begin{equation}
    K=K(\lambda,\mu,t)=\frac{\lambda (\lambda -1)  (\lambda -t)}{{t(t-1) }} \left[\mu ^2-\left(\frac{\theta _0}{\lambda }+\frac{\theta _1}{\lambda -1}+\frac{\theta _t-1}{\lambda -t}\right)\mu  +\frac{\kappa _2\left(\kappa _1-1\right) }{\lambda(\lambda -1)  }\right].
\end{equation}
One notes that $q_0$ drops out of the Hamiltonian.

We can now write down the Hamilton equations:
\begin{align}\label{eqLK}
    &\frac{d\lambda}{dt}=\frac{\partial K}{\partial \mu}=\frac{\lambda (\lambda -1)  (\lambda -t)}{t(t-1)} \left(2 \mu-\frac{\theta _0}{\lambda }-\frac{\theta _1}{\lambda -1} -\frac{\theta _t-1}{\lambda -t}\right),
    \\\nonumber
    &\frac{d\mu}{dt}=-\frac{\partial K}{\partial \lambda}=-K\bigg(\frac{1 }{\lambda  }+\frac{1 }{\lambda -1 }+\frac{1}{\lambda-t }\bigg)
    \\\label{eqMuK}
&\qquad\qquad\qquad\,\,+\frac{\lambda(\lambda -1)   (\lambda -t) }{t(t-1)}\left[\frac{\kappa _2 \left(\kappa _1-1\right) (2 \lambda -1)}{\lambda ^2(\lambda -1)^2 }-\mu  \left(\frac{\theta _0}{\lambda ^2}+\frac{\theta _1}{(\lambda -1)^2}+\frac{\theta _t-1}{(\lambda -t)^2}\right)\right].
\end{align}
This system is equivalent to the second order non-linear ordinary differential Painlev\'e VI equation for $\lambda(t)$ or $\mu(t)$. In order to derive this equation for $\lambda(t)$ for example, one eliminates $\mu(t)$ from Eq. (\ref{eqLK}):
\begin{equation}
    \mu(t)=\frac{t (t-1)}{2 \lambda (\lambda-1)(\lambda-t)}\lambda'(t)+\frac{1}{2}\bigg(\frac{\theta_0}{\lambda}+\frac{\theta_1}{\lambda-1}+\frac{\theta_t-1}{\lambda-t}\bigg).
\end{equation}
Now we take the derivative with respect to $t$:
\begin{align}
  \nonumber  \mu'(t)&=\frac{1}{2} \left(\frac{\lambda -\theta _0}{\lambda ^2}-\frac{\theta _1}{(\lambda -1)^2}+\frac{1}{1-\lambda }+\frac{2-\theta _t}{(t-\lambda )^2}\right) \lambda '(t)
    \\
    &+\frac{1}{2}\bigg(\frac{1-t}{ \lambda ^2}+\frac{t}{(\lambda -1)^2}-\frac{1}{(\lambda -t)^2}\bigg) \big(\lambda '(t)\big)^2+\frac{t(t-1) }{2 \lambda(\lambda -1)   (\lambda -t)}\lambda ''(t)
    +\frac{\theta _t-1}{2 (\lambda -t)^2}.
\end{align}
Equating this to the right hand side of Eq. (\ref{eqMuK}) one finds the Painlev\'e VI equation:
\begin{align}
   \nonumber \lambda''(t)&=\frac{1}{2} \bigg(\frac{1}{\lambda}+\frac{1}{\lambda-1}+\frac{1}{\lambda-t}\bigg) \lambda^{\prime\, 2}-\bigg(\frac{1}{t}+\frac{1}{t-1}+\frac{1}{\lambda-t}\bigg)\lambda'
    \\
   & +\frac{\lambda (\lambda-1)(\lambda-t)}{2 t^2 (t-1)^2} \bigg((\theta_\infty-1)^2- \theta_0^2\frac{t}{\lambda^2} -\theta_1^2\frac{t-1}{(\lambda-1)^2}-(\theta_t^2-1) \frac{t(t-1)}{(\lambda-t)^2}\bigg),
\end{align}
where $\theta_\infty=\kappa_1-\kappa_2=\alpha-\beta$. We can compare this with the standard form of the PVI,  
\begin{align}\label{StandardPVI}
   \nonumber \lambda''(t)&=\frac{1}{2} \bigg(\frac{1}{\lambda}+\frac{1}{\lambda-1}+\frac{1}{\lambda-t}\bigg) \lambda^{\prime\, 2}-\bigg(\frac{1}{t}+\frac{1}{t-1}+\frac{1}{\lambda-t}\bigg)\lambda'
    \\
   & +\frac{\lambda (\lambda-1)(\lambda-t)}{ t^2 (t-1)^2} \bigg(\tilde\alpha+\tilde\beta\frac{ \,t}{\lambda^2}+\tilde\gamma \frac{ t-1}{(\lambda-1)^2}+\tilde\delta\frac{t(t-1)}{(\lambda-t)^2}\bigg),
\end{align}
which yields the following coefficients:
\begin{equation}\label{eqPVIcoefs}
\tilde\alpha=\frac{(\theta_{\infty}-1)^2}{2}, \quad \tilde\beta=-\frac{\theta_0^2}{2}, \quad \tilde\gamma=-\frac{\theta_1^2}{2}, \quad \tilde\delta=-\frac{\theta_t^2-1}{2}.  
\end{equation}

The derivation of PVI was general. In particular if we consider the scalar dynamics in $Y^{p,q}$ the characteristic exponents $\theta_{0,1,t,\infty}$ assume the form given in Eqs. (\ref{eqSomeThetaEqs12})-(\ref{eqSomeThetaEqs13}).

\subsection{Fuchsian equations with three singularities and Painlev\'e V} \label{sectPV}

In general the Fuchsian equation leading to PV is going to be different from the general Heun equation. Therefore, in this subsection we will use $\bar\theta_i=\bar\theta_0,\bar\theta_t$ for the characteristic exponents to discern them from the Heun's ones $\theta_i=\theta_0,\theta_1,\theta_t$.   
In the Painlev\'e V case one still has Eq. (\ref{eqDeformedHeun}). The difference now is the ansatz for the apparent singularity $\lambda$ \cite{cavalcante:2019}:
\begin{equation}\label{eqA12aV}
    A_{12}(z)=\frac{\zeta (z-\lambda)}{z (z-t)}=\frac{A_{0,12}}{z}+\frac{A_{t,12}}{z-t}, \quad \zeta=const.
\end{equation}
Finding the common denominator in the r.h.s we can compare the coefficients in powers of $z$ to find:
\begin{equation} \label{klamform}
    \zeta=A_{0,12}+A_{t,12},\quad \lambda=\frac{A_{0,12}}{A_{0,12}+A_{t,12}} t.
\end{equation}

We prefer to work with the components of the matrices $A_i$. In this case one has
\begin{equation}
    A(z)=\sum\limits_{i=0}^t \frac{A_{i}}{z-z_i}+A_\infty=\frac{A_0}{z}+\frac{A_t}{z-t} + A_\infty,
\end{equation}
where the components of the matrix $A$ are given by:
\begin{align}
   & A_{11}(z)=\frac{A_{0,11}}{z}+\frac{A_{t,11}}{z-t}+A_{\infty,11},\qquad
   A_{12}(z)=\frac{A_{0,12}}{z}+\frac{A_{t,12}}{z-t}
   \\
& A_{21}(z)=\frac{A_{0,21}}{z}+\frac{A_{t,21}}{z-t},\qquad\qquad\qquad\!\!
A_{22}(z)=\frac{A_{0,22}}{z}+\frac{A_{t,22}}{z-t}+A_{\infty,22}.
\end{align}
In the expressions above we assumed a diagonal form of $A_\infty$, i.e. $A_{\infty,12}=A_{\infty,21}=0$.
Upon substitution of these components and the ansatz (\ref{eqA12aV}) of $A_{12}(z)$ in (\ref{eqPQz}), one finds
\begin{align} \label{PPV}
    &P(z)=-\bigg(\tr A_\infty+\frac{\tr A_0-1}{z}+\frac{\tr A_t-1}{z-t}+\frac{1}{z-\lambda}\bigg),
\\
 \label{QPV}
& Q(z) = \det A_\infty + \frac{\det A_0}{z^2} + \frac{\det A_t}{(z-t)^2} + \frac{c_0}{z} - \frac{c_t}{z-t} + \frac{\tilde\mu}{z-\lambda} ,
\end{align}
where 
\begin{equation}
       \tilde \mu=A_{\infty,11}+\frac{1}{\lambda} A_{0,11}+\frac{1}{z-t} A_{t,11},
\end{equation}
and $c_0$ and $c_t$ given by 
\begin{align}
    c_0=\tr A_\infty \tr A_0-\tr(A_\infty A_0)+\frac{ \tr(A_0 A_t)}{t}-\frac{ \tr A_0 \tr A_t}{t} -A_{\infty,11}-\frac{A_{0,11}}{\lambda} +\frac{A_{0,11}+A_{t,11}}{t},
\end{align}
\begin{align}
    c_t=\tr A_\infty \tr A_t-\tr(A_\infty A_t)-\frac{ \tr(A_0 A_t)}{t}+\frac{ \tr A_0 \tr A_t}{t} -A_{\infty,11}-\frac{A_{t,11}}{\lambda-t} -\frac{A_{0,11}+A_{t,11}}{t}.
\end{align}
The variable $\mu$ follows from the residue of $Q(z)$ at $z=\lambda$:
\begin{eqnarray}\label{muPV}
    \mu=\mathop{\rm{Res}}_{z=\lambda}Q(z)=A_{\infty,11}+\frac{1}{\lambda} A_{0,11}+\frac{1}{\lambda-t} A_{t,11}.
\end{eqnarray}

Furthermore, since our singularities $0, t$ are regular their connection matrices can be diagonalized and thus we can choose a gauge similar to the one in Eq. \eqref{eqAparam}:
\begin{equation} \label{gaugePV}
\tr A_i = \bar\theta_i, \; \; \det A_i = 0 \;\; i \in \{ 0, t\},\quad A_\infty=\frac{1}{2}\sigma^3=\frac{1}{2} \left(
\begin{array}{cc}
 1 & 0 \\
 0 & -1 \\
\end{array}
\right).
\end{equation}
In this case we need to make an additional choice of gauge in order to fix the form of all matrix elements of our connections. Let us multiply to the left both sides of the `residue' relation $A_\infty=-A_0-A_t$ by $A_\infty$ and take the trace. Hence, one finds \cite{cavalcante:2019}: 
\begin{equation} \label{mystery}
    \tr(A_\infty (A_0 + A_t)) = -\frac{\bar\theta_\infty}{2},
\end{equation}
where $\bar\theta_\infty$ is the difference between characteristic exponents at infinity. Applying these assumptions to equation \eqref{QPV} and using the general form \eqref{resH} for the Hamiltonian, we obtain:

\begin{align} \label{HPV}
    \nonumber K& = \frac{1}{2} (1 + A_{t,1 1}  - A_{t,2 2} ) + \frac{ A_{t,11} } {\lambda - t}+
    \\ 
    &+ \frac{ A_{0,11} + A_{t,1 1} - A_{0,22} A_{t,1 1} + A_{0,2 1}  A_{t,12} + A_{0,1 2} A_{t,2 1} - A_{0,1 1} A_{t,2 2}}{t}.
\end{align} 

In order to express the Hamiltonian in terms of $K=K(\lambda, \mu,t)$, we have to eliminate the matrix elements one by one. For this purpose we first solve Eq. (\ref{muPV}) for $A_{0,11}$ and use $\tr A_0=A_{0,11}+A_{0,22}=\bar\theta_0$ to express $A_{0,22}$:
\begin{equation}
    A_{0,11}=\lambda \bigg(\mu-\frac{1}{2}-\frac{A_{t,11}}{\lambda-t}\bigg), \quad A_{0,22}=\bar\theta_0-A_{0,11}=\bar\theta_0-\lambda \bigg(\mu-\frac{1}{2}-\frac{A_{t,11}}{\lambda-t}\bigg),
\end{equation}
Next we insert these two expressions in $\det A_0=0=A_{0,11} A_{0,22}-A_{0,12} A_{0,21}$ to find $A_{0,21}$:
\begin{equation}
A_{0,21}=\frac{\lambda}{A_{0,12}}  \bigg(\mu-\frac{1}{2}-\frac{A_{t,11}}{\lambda-t}\bigg)\bigg[\bar\theta_0-\lambda \bigg(\mu-\frac{1}{2}-\frac{A_{t,11}}{\lambda-t}\bigg)\bigg].
\end{equation}
We now solve $\tr A_t=A_{t,11}+A_{t,22}=\bar\theta_t$ to express $A_{t,22}$:
\begin{equation}
    A_{t,22}=\bar\theta_t-A_{t,11}.
\end{equation}
Then we insert the latter expression into $\det A_t=0=A_{t,11} A_{t,22}-A_{t,12} A_{t,21}$ to solve for $A_{t,21}$:
\begin{equation}
    A_{t,21}=\frac{A_{t,11}(\bar\theta_t-A_{t,11})}{A_{t,12}}.
\end{equation}
Now we need Eq. (\ref{mystery}), which explicitly yields:
\begin{equation}
    (A_{0,11}+A_{t,11}) -(A_{0,22}+A_{t,22}) =-\bar\theta_\infty.
\end{equation}
We can use it to express $A_{t,11}$ by inserting the expressions for $A_{0,11}$, $A_{0,22}$ and $A_{t,22}$:
\begin{equation}
A_{t,11}=-\frac{\lambda-t}{2 t} \big(\bar\theta_0+\bar\theta_t-\bar\theta_\infty+\lambda(1-2 \mu)\big).
\end{equation}
The last equation we need is (\ref{klamform}), which we can solve with respect to $A_{t,12}$:
\begin{equation}
     A_{t,1 2} =  A_{0,1 2}\bigg(\frac{t}{\lambda} - 1\bigg).
\end{equation}
Finally, we have everything to   to obtain the explicit form of the Hamiltonian by inserting the expressions for $A_{0,11}$, $A_{0,22}$, $A_{0,21}$, $A_{t,22}$, $A_{t,21}$, $A_{t,11}$ and $A_{t,12}$ in that order to find:
\begin{equation} \label{HPVf}
   K = \frac{\lambda (\lambda -t ) \mu^2}{t} + \frac{\lambda (1- \bar\theta_t ) -\bar\theta_0 (\lambda-t)}{t} \mu - \frac{(\lambda-t)}{4t} \big( \lambda+2(1+ \bar\theta_\infty)\big).
\end{equation}
One notes that the matrix element $A_{0,12}$ drops out of the Hamiltonian, thus we have an explicit function of $K=K(\mu,\lambda,t)$.

We can use this form of the Hamiltonian to find the equations governing our flow. Employing a similar technique to the one used in Section \ref{sectPVI}, to obtain the following second order equation:
\begin{align}
    \lambda'' &= \frac{1}{2}\bigg(\frac{1}{\lambda} + \frac{1}{\lambda - t}\bigg)\lambda^{\prime \,2} -\bigg(\frac{1}{t}+\frac{1}{\lambda-t}\bigg) \lambda' 
    +\frac{\lambda  \bar\theta _{\infty } (t-\lambda )}{t^2}+\frac{\lambda  \bar\theta _t^2}{2 t^2-2 \lambda  t}+\frac{\bar\theta _0^2 (\lambda -t)}{2 \lambda  t}
    \\
    &+\frac{\lambda ^4}{t^2 (\lambda -t)}+\frac{\lambda ^3 (5 t-2)}{2 t^2 (t-\lambda )}-\frac{2 \lambda ^2 (t-1)}{t (t-\lambda )}+\frac{\lambda  ((t-2) t-1)}{2 t (t-\lambda )},
\end{align}
where primes are equivalent to partial derivatives with respect to $t$. This is Painlev\'e V equation, but not in a standard form. We can get the standard equation by first considering the transformation $\lambda \rightarrow t \lambda$ \cite{slavyanov:2018}. Upon transforming we get the following form of the isomonodromic flow:
\begin{align}
    \lambda'' &= \frac{1}{2} \bigg( \frac{1}{\lambda} + \frac{1}{\lambda - 1} \bigg) \lambda^{\prime\, 2} - \frac{1}{t} \lambda' -\frac{\lambda  \bar\theta _t^2}{2 (\lambda -1) t^2}+\frac{\bar\theta _0^2 (\lambda -1)}{2 \lambda  t^2}-\frac{(\lambda -1) \lambda  \bar\theta _{\infty }}{t}
    \\
    &+ \frac{\lambda ^4}{\lambda -1}+\frac{\lambda ^3 (2-5 t)}{2 (\lambda -1) t}+\frac{2 \lambda ^2 (t-1)}{(\lambda -1) t}-\frac{\lambda  (t-2)}{2 (\lambda -1) t}.
\end{align}
Next step is to take into account that we can swap\footnote{This follows from the fact that the singularities of the Heun equation are invariant under the action of the $D_4$ Coexeter group.} singularities freely and thus consider the transformation of the form $\lambda \rightarrow \frac{\lambda}{\lambda - 1}$, which is equivalent to exchanging the positions of the singularities at 1 and at $\infty$ \cite{slavyanov:2000}. After transforming the equation, we get:
\begin{align}
     \lambda'' &= \bigg(\frac{1}{2 \lambda} + \frac{1}{\lambda - 1}\bigg)\lambda^{\prime\,2} - \frac{1}{t} \lambda' 
    + \frac{(\lambda -1)^2 }{t^2} \left(\frac{ \bar\theta _t^2}{2}\lambda -\frac{\bar\theta _0^2}{2 \lambda }\right)+(\bar\theta _{\infty }-1)\frac{\lambda  }{t}-\frac{1}{2}\lambda \frac{ \lambda +1 }{\lambda -1}.
\end{align}
We can now compare to the fifth Painlev\'e equation in its standard form:
\begin{equation} \label{PVstandard}
    \lambda'' = \bigg(\frac{1}{2 \lambda} + \frac{1}{\lambda - 1}\bigg)\lambda^{\prime\,2} - \frac{1}{t} \lambda' + \frac{(\lambda - 1)^2}{t^2} \bigg(\bar\alpha \lambda - \frac{\bar\beta}{\lambda}\bigg) + \bar\gamma \frac{\lambda}{t} + \bar\delta \lambda \frac{\lambda + 1}{\lambda - 1}
\end{equation}
to determine the parameters $\tilde\alpha, \tilde\beta, \tilde\gamma$ and  $\tilde\delta$  \cite{slavyanov:2000} such as:
\begin{equation}\label{tilde_coeff}
    \bar\alpha=\frac{\bar\theta_t^2}{2},\quad \bar\beta=\frac{\bar\theta_0^2}{2},\quad \bar\gamma=\bar\theta_\infty-1,\quad \bar\delta =- \frac{1}{2}.
\end{equation}

The derivation of PV was general. Now the values of the exponents $\bar \theta_{0,t,\infty}$ will depend on how we approach the limit $q\to p$.

\begin{comment}
In order to show that the two forms are equivalent, we need to take into account that we can swap\footnote{This follows from the fact that the singularities of the Heun equation are invariant under the under the action of the $D_4$ Coexeter group.} singularities freely and thus consider the transformation of the form $\lambda \rightarrow \frac{\lambda}{\lambda - 1}$, which is equivalent to exchanging the positions of the singularities at 1 and at $\infty$ \cite{slavyanov:2000}. After transforming the equation, we get:
\begin{equation}
    \lambda'' = \frac{1}{2} \bigg(\frac{1}{\lambda} - \frac{1}{\lambda - 1}\bigg) \lambda^{\prime \,2} - \frac{1}{t} \lambda' + \frac{1}{t^2} \bigg(\frac{\beta}{\lambda} + \frac{\beta - (\alpha + \beta)\lambda}{\lambda - 1}\bigg) - \frac{\gamma}{t} \lambda (\lambda - 1) - \delta \lambda (\lambda - 1)(2 \lambda - 1).
\end{equation}
Therefore for an appropriate choice of parameters $\alpha, \beta, \gamma, \delta$, the equations are equivalent.
\end{comment}

\subsection{From PVI to PV via coalescence of singularities}

The fifth Painlev\'e equation can be obtained by the confluent limit on Painlev\'e VI,
where the limit is obtained by the confluence of two singular points and scaling transformations on the PVI parameters. Replacing the parameters in the standard PVI (\ref{StandardPVI}) by
\begin{equation}
    (t,\lambda,\tilde\alpha,\tilde\beta,\tilde\gamma,\tilde\delta) \to (1+\epsilon t,\lambda,\hat\alpha,\hat\beta,\epsilon^{-1}\hat\gamma-\epsilon^{-2}\hat\delta,\epsilon^{-2}\hat\delta)
\end{equation}
and then taking the limit $\epsilon\to 0$, we obtain the standard PV equation:
\begin{align}\label{eqPVafterCoalesence}
\lambda'' = \bigg(\frac{1}{2 \lambda} + \frac{1}{\lambda - 1}\bigg)\lambda^{\prime\,2} - \frac{1}{t} \lambda' + \frac{(\lambda - 1)^2}{t^2} \bigg(\hat\alpha \lambda - \frac{\hat\beta}{\lambda}\bigg) + \hat\gamma \frac{\lambda}{t} + \hat\delta \lambda \frac{\lambda + 1}{\lambda - 1},
\end{align}
where the coefficients $\hat\alpha,\hat\beta,\hat\gamma,\hat\delta$ coincides with $\tilde\alpha,\tilde \beta,\tilde\gamma,\tilde \delta$ from (\ref{eqPVIcoefs}).

\section{Applications to the Y$^{p,q}$ and its confluent limits} \label{apps}

Let us make a short summary of the results from the previous sections. The monodromy  data (Eqs.(\ref{eqSomeThetaEqs12})-(\ref{eqSomeThetaEqs13})) for the general Heun equation \textbf{GHE} in $Y^{p,q}$ is given by
\begin{align} \label{gheGHE}
\theta _0 =-2|\alpha_1|,
\quad\theta _1=-2|\alpha_2|,
\quad
 \theta _t=-2|\alpha_3|,
\quad \theta_\infty=-2 (\kappa+1),
\end{align}
where $\alpha_{1,2,3}$ and $\kappa$ have been defined in Eqs.(\ref{eqAlpha1})-(\ref{eqAlpha3}) and (\ref{eqenergyKappa}). The isomonodromic deformation of \textbf{GHE} leads to the Painlev\'e VI (\textbf{PVI}) equation (\ref{StandardPVI}) with coefficients (\ref{eqPVIcoefs}):
\begin{equation}\label{eqAlphaTildas}
\tilde\alpha=\frac{(\theta_{\infty}-1)^2}{2}, \quad \tilde\beta=-\frac{\theta_0^2}{2}, \quad \tilde\gamma=-\frac{\theta_1^2}{2}, \quad \tilde\delta=-\frac{\theta_t^2-1}{2}.  
\end{equation}
The limit $q\to p$ in \textbf{PVI} leads to a new \textbf{PVI$'$} equation with coefficients:
\begin{align}\label{eqAplhaPrimes1}
\nonumber &\alpha'=\frac{\big(\theta^{(q\to p)}_\infty-1\big)^2}{2}=\frac{(2\kappa+3)^2}{2},\quad  \beta'=-\frac{\big(\theta_0^{(q\to p)}\big)^2}{2}=-\frac{(N_\psi-p N_\alpha)^2}{2},
    \\
&\gamma'=-\frac{\big(\theta_1^{(q\to p)}\big)^2}{2}=-\frac{N_\psi^2}{2},\quad \delta'=-\frac{\big(\theta_t^{(q\to p)}\big)^2-1}{2}=\frac{1-N_\psi^2}{2}.
\end{align}
After a coalescence procedure in \textbf{PVI$'$} one obtains \textbf{PV$'$} equation with the same coefficients as in (\ref{eqAplhaPrimes1}). On the other hand, after a coalescence procedure in \textbf{PVI} one obtains \textbf{PV} equation with the same coefficients as in (\ref{eqAlphaTildas}). Taking the limit $q \to p$ in \textbf{PV}, one obtains \textbf{PV$''$} with the same coefficients as in \textbf{PV$'$} \eqref{eqAplhaPrimes1}.

Before we consider the other limits of the geometry, examined in previous sections, let us make some remarks about the corresponding Painlev\'e equations. The subsequent limits of Y$^{p,q}$ have either resonant or irregular singular points, whose cases have not been considered in Section \ref{minv}. The latter is due to the fact that the isomonodromic method requires an extension including the Stokes phenomenon, which we do not consider here. Nevertheless, since the respective equations of motion have three singularities, we know that the corresponding isomonodromic flow equations will be of the Painlev\'e V class.

\begin{figure}[H]
\begin{center}
\begin{tikzpicture}[scale = 1.3]
%above branch
\node[draw] (Ypq) at (0,0) {$Y^{p,q}$};
\node[draw] (HE)  at (2.5,1) { GHE };
\node[draw] (PVI) at (2.5,3) { PVI};
\node[draw] (PV2) at (5.5,2) { PV};
\node[draw] (PV3) at (8,2) { PV$^{''}$};

\node[draw] (PVI2) at (5.5,4) { PVI$^{'}$};
\node[draw] (PV)   at (8,4) { PV$'$};

%middle branch
\node[draw] (GHG2) at (5.5, 1) { HypG };
\node[draw] (PV1) at (8, 0) { PV$^{''' *}$};

%1,0 branch
\node[draw] (T11) at (2.5, -1.5) {Y$^{1,0} \cong T^{1,1} / \mathbb{Z}_2$};
\node[draw] (HG10) at (5.5, -1.5) {HypG$_2$};
\node[draw] (PVt) at (8, -1.5) { $\tilde{PV}^*$};

\draw[thick, ->] (Ypq) -- (T11);
\draw[thick, ->] (T11) -- (HG10) node[midway,above] {EoM};
\draw[thick, ->] (HG10) -- (PVt) node[midway,above] {isom.};

%inf, q branch
\node[draw] (Yinf) at (1.5, -3) {Y$^{\infty, q}$};
\node[draw] (CHinf) at (5.5, -3) {CHE};
\node[draw] (PVb) at (8, -3) { $\overbar{PV}^*$};

\draw[thick, ->] (Ypq) -- (Yinf) node[midway,below] { \rotatebox{-65}{$p \to \infty$}};
\draw[thick, ->] (Yinf) -- (CHinf) node[midway,above] {EoM};
\draw[thick, ->] (CHinf) -- (PVb) node[midway,above] {isom.};

%above branch arrows
\draw[thick, ->] (Ypq) -- (HE);  
\draw (1,0.65) node{\rotatebox{22}{EoM}};
\draw[thick, ->] (HE) -- (PVI);
\draw (2.27,2) node{ \rotatebox{90}{isom.}};
\draw[thick, ->] (PVI) -- (PV2);
\draw (4,2.3) node { \rotatebox{-17.5}{coal.}};
\draw[thick, ->] (PVI2) -- (PV) node[midway,above] { coal.};

\draw[thick, ->] (PV2) -- (PV3);
\draw (6.7,2.19) node{ $q \rightarrow p$};
\draw[thick, ->] (PVI) -- (PVI2);
\draw (4,3.7) node { \rotatebox{17.5}{$q \rightarrow p$}};
\draw[thick, <->] (PV) -- (PV3);
\draw (8.2,3) node{\rotatebox{90}{equiv.}};

\draw[thick, ->] (HE) -- (GHG2);
\draw (4,1.19) node { $q \rightarrow p$};

\draw[thick, ->] (GHG2) -- (PV1);
\draw (6.85,0.68) node {\rotatebox{-22}{ isom.}};

%\draw [dashed] (8,1.7) -- (8,0.25);
%\draw (8.2,1) node{\rotatebox{90}{\small{not equiv.}}};
\end{tikzpicture}
\caption{A schematic representation of the results obtained in this work. EoM specifies the scalar field equation of motion. Isom. indicates that isomonodromic deformation has been performed on the corresponding EoM. Coal. states a coalescence procedure between two Painlev\'e equations. Equiv. shows equivalence between the corresponding equations.  The explicit form of all equations marked with $*$ has not been found due to limitations of the derived isomonodromic deformation method. The latter is related to two different key points, namely the presence of resonant or irregular singularities. }\label{figScheme2}
\end{center}
\end{figure}
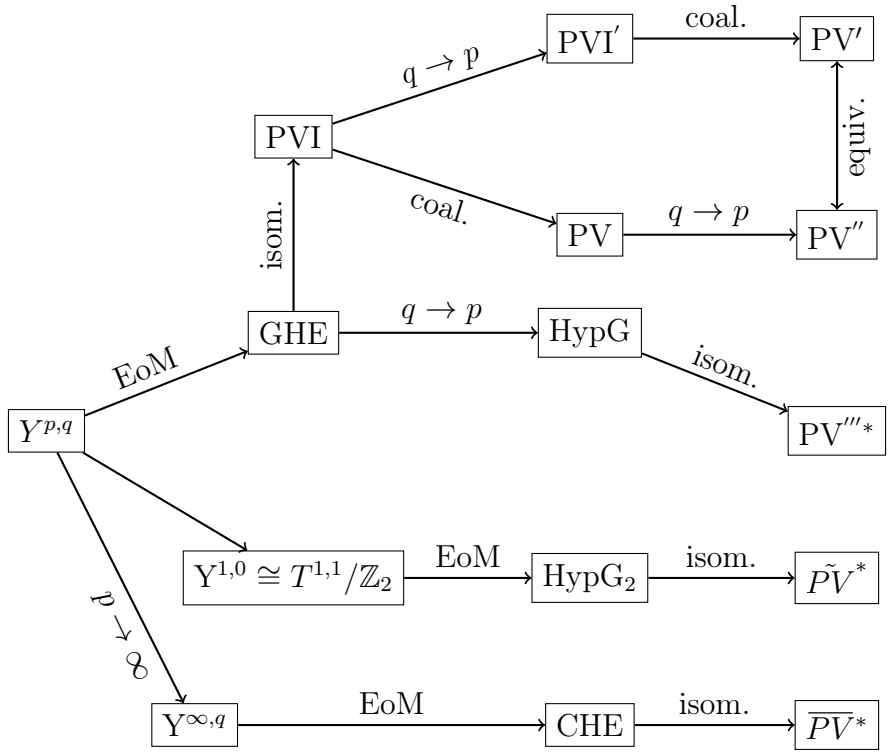

Let us go back again to \textbf{GHE} and take the limit $q \to p$. In this case one produces a hypergeometric equation  \textbf{HypG} with coefficients as in \eqref{ppHypTh}. The latter equation has cases with resonant singularities. The treatment of these singularities requires the splitting of the monodromy space in sectors, leading to a generalization of the isomonodromic method. This will be a subject of a separate work. As noted earlier, the corresponding Painlev\'e equation to \textbf{HypG} will fall under the PV class. In the diagram \ref{figScheme2}, it is noted as \textbf{PV$'''$}.

Looking back at the original geometry we can consider the $Y^{1,0} \cong T^{1,1} / \mathbb{Z}_2$ limit. As shown in \cite{arnaudov2011pulsating,Arnaudov:2010dk}, the equations of motion of these geometries also lead to a hypergeometric equation \textbf{HypG$_2$}. The explicit form of the parameters can be found in equation \eqref{thY10}. A consequent generalized isomonodromic deformation will lead to $\tilde{PV}$ equation.

Finally, taking $p \to \infty$ one again ends up with a confluent Heun equation \textbf{CHE}. The form of its coefficients is given in equation \eqref{thYinf}. It is well known that the confluent Heun equation has two regular and one irregular singularities. The irregular singularities complicates the structure of the monodromy data by introducing Stokes sectors. However, it is certain that the corresponding Painlev\'e equation will be of the PV class.

\section{Conclusion} \label{conc}

The description of Super Yang-Mills theories in the context of holography has been a fruitful area of investigation. Finding gravity duals of such theories and vice-versa is a non-trivial task. One of the recent advances in this area is the discovery of the infinite sets of Sasaki-Einstein geometries Y$^{p,q}$ and L$^{p,q,r}$ \cite{Benvenuti:2004dy, Cvetic:2005ft}. The duals of these are known to be quiver gauge theories, whose complete properties are an active area of research. The latter naturally arise in the construction of gauge theories from D-branes in string theory. Furthermore, Sasaki-Einstein spaces appear in the context of emergent spacetime, where the classical geometry of the bulk is hypothesized to emerge from the underlying quantum dynamics of a corresponding gauge theory \cite{berman2021emergent, berman2022emergent}.  

Motivated by the progress of this topic, we focused on the dynamics of scalar fields in Sasaki-Einstein Y$^{p,q}$ backgrounds. More specifically, we implemented the theory of isomonodromic deformations to the arising equations of motion in Y$^{p,q}$. As a first step, our  goal is to obtain the corresponding non-linear Painlev\'e class equations and classify the corresponding limiting cases in Y$^{p,q}$. The limits of the original geometry were chosen to induce a confluence of singularities of the original geometry. Such cases are inherently interesting from the dual theory point of view, as they result in a phase transition of the original theory. In essence, each quiver/gauge group in the CFT corresponds a toric base in the toric diagram of the corresponding geometry. Therefore, to each limit of our geometry we can associate a different quiver theory with a smaller number of vertices. Thus, we can infer differences between the dual CFTs by comparing their monodromy data and corresponding Painlev\'e classes.

We first considered the standard Y$^{p,q}$ geometry, whose equations of motion for scalar fields reduce to a general Heun equation \cite{Kihara:2005nt}. As an extension, we have shown that the isomonodromic method yields the Painlev\'e VI equation \eqref{eqAlphaTildas}, which has not been done previously for Sasaki-Einstein backgrounds.

In the context of this result, we performed various limits of the original background, which result in coalescence of singularity and by extension simplifications of the toric diagrams. The first limit is $q \to p$, which we show to reduce the original equations of motion to a hypergeometric one. The resulting equation has an interesting property, where some of the singular points can oscillate between resonant and non-resonant, depending on the values of the quantum numbers, which can be seen by \eqref{ppHypTh}. We obtain a similar result for the T$^{1,1} / \mathbb{Z}_2$ geometry, which is known to be isomorphic to Y$^{1,0}$ \cite{Gauntlett:2004yd}. For this limit, the EoMs also reduce to a hypergeometric equation \eqref{hgT11} and its resonant points can express a similar oscillation to the Y$^{p,p}$ case. It is well known that in both of the limiting cases the equations of motion correspond to a Painlev\'e V. Furthermore these two cases (and Y$^{2,2}$) are the only Y$^{p,q}$ geometries admitting supersymmetric deformations, which forces one to conjecture a connection between the monodromy data of the original geometry and the properties of its dual conformal theory.

The final confluent limit considered in this work is the Y$^{\infty,q}$. For this case we have shown that the original equations of motion reduce to a confluent Heun equation. This equation again corresponds to a Painlev\'e V class equation, however the explicit form is hard to obtain since it has an irregular singularity at $\infty$. This implies that the explicit form of the Painlev\'e equation will be affected by the presence of Stokes multipliers. This will have an effect on the topology of the Painlev\'e monodromy manifolds associated to the monodromy data of the problem \cite{chekhov:2017}. This result, combined with the results for the other confluent limits, allow us to strengthen the conjecture that there is a connection between the topological properties of the monodromy manifolds of an AdS space and the properties of its dual superconformal theory.

\begin{comment}
The form of the EoM when $p > q$ is a general Heun equation with four regular singularities. The latter is important due to the fact that the isomonodromic method, derived in Section \ref{minv}, is directly applicable for equations with regular non-resonant singularities. Hence we were able to derive the corresponding non-linear Painlev\'e VI equation, which has not been done previously for Sasaki-Einstein backgrounds. This result is important due to the recent advances to the theory of Frobenius manifolds related to the monodromy data of the problem. The latter is expected to give an important new perspective to the algebraic and geometric structure of Sasaki-Einstein spaces and other holographic backgrounds.

In this context, we also considered various interesting limits of the original Y$^{p,q}$ background. The first limit is given by $q \to p$, which we show reduces the original equation of motion to a hypergeometric one with varying number of resonant singularities, depending on the chosen quantum numbers. The latter complicates the structure of the monodromy data and requires the extension of the original isomonodromic method. Another limit was Y$^{1,0}$, which is known to be isomorphic to $T^{1,1} / \mathbb{Z}_2$. Analogously, we encounter a similar situation with unavoidable resonances. The final limit was Y$^{\infty, q}$, which was complicated by the occurrence of irregular singularities, which once again requires a further extension of the isomonodromic method.
\end{comment}

It would be interesting to consider the generalization of the isomonodromic method for resonant and/or irregular singular points. By extension, this would allow us to extract more data about the topology of the confluent limit and their geometries, relating it to the properties of the dual theories. Another direction is to generalize the considerations to the L$^{p,q,r}$ family of Sasaki-Einstein backgrounds. These subjects are an active area of investigation in the theory of differential equations and algebraic topology, potentially leading to novel results with important physical applications. We hope to revisit this problem in subsequent works.
\section*{Acknowledgements}

The authors would like to thank Goran Djordjevic,  Dragoljub Dimitrijevic and the SEENET-MTP for the warm hospitality during BPU11 and satellite events. We also gratefully acknowledge the support of the Simons Foundation and the International Center for Mathematical Sciences in Sofia for the various annual scientific events. H. D. and M. R. thankfully acknowledge the support by the program “JINR—Bulgaria” of the Bulgarian Nuclear Regulatory Agency. M. R. is thankful for the support by the Bulgarian national program “Young Scientists and Postdoctoral Research Fellows.” RD-22-1719. T. V. is grateful to the SEENET-MTP - ICTP Program NT03. R. R. was partially supported by the NSF grant H28/5. 

\appendix

\section{General Heun equation in normal and canonical forms}

The six parametric normal form of the general Heun equation (GHE) is\cite{cht:2004}:
\begin{equation}\label{eqHeunNormalForm}
H''(z)-\left(\frac{A_1}{z}+\frac{A_2}{z-1}+\frac{A_3}{z-\tau}+\frac{A_4}{z^2}+\frac{A_5}{(z-1)^2}+\frac{A_6}{(z-\tau)^2}\right) H(z)=0.
\end{equation}
The traditional (canonical) form of Heun's equation is given by 
\begin{equation}\label{eqStandardHeun}
h''(z)+\left(\frac{\gamma }{z}+\frac{\delta }{z-1}+\frac{\epsilon }{z-\tau}\right)h'(z)+\frac{\alpha  \beta  z-k}{z (z-1) (z-\tau)} h(z)=0,
\end{equation}
together with the Fuchsian relation 
\begin{equation}\label{Fuch_rel}
    \alpha +\beta +1=\gamma +\delta +\epsilon, \quad  \tau\neq 0,1.
\end{equation}
The two forms are related by the following transformation:
\begin{equation}\label{eqTransformToStandard}
H(z)=z^{\gamma/2} (z-1)^{\delta/2} (z-\tau)^{\epsilon/2}h(z),
\end{equation}
together with the relations:
\begin{align}\label{eqAsWithHeunP}
\nonumber &A_1=\frac{k}{\tau}-\frac{\gamma  \epsilon }{2 \tau}-\frac{\gamma  \delta }{2},\quad A_2=\frac{\gamma  \delta }{2}-\frac{\delta  \epsilon }{2 (\tau-1)}-\frac{k-\alpha  \beta }{\tau-1},
\quad A_3=\frac{\gamma  \epsilon }{2 \tau}+\frac{\delta  \epsilon }{2 (\tau-1)}-\frac{\tau \alpha  \beta -k}{\tau (\tau-1)},
\\
&A_4=\frac{\gamma }{2}  \left(\frac{\gamma }{2}-1\right),\quad A_5=\frac{\delta }{2}  \left(\frac{\delta }{2}-1\right),\quad A_6=\frac{\epsilon }{2}  \left(\frac{\epsilon }{2}-1\right).
\end{align}
In order to get these coefficients one has to convert (\ref{eqStandardHeun}) to normal form by taking out the first derivative and compare coefficients in front of the powers of $z$. Any second order equation in the form $\varphi''(z)+\mathcal{P}(z) \varphi'(z)+\mathcal{Q}(z) \varphi(z)=0$ can be brought to normal form, $f''(z)-V(z)f(z)=0$, by $V(z)=-\mathcal{Q}(z)+\frac{1}{2} \mathcal{P}(z)'+\frac{1}{4} \mathcal{P}^2(z)$, where $\ln \varphi(z)=\ln f(z)-\frac{1}{2}\int \mathcal{P}(z) dz$.

Another canonical form of Heun equation, which is more useful to us, is given by
\begin{equation}\label{eqHeunThetaForm}
    y''(z)+\bigg(\frac{1-\theta_0}{z}+\frac{1-\theta_1}{z-1}+\frac{1-\theta_{t}}{z-t}\bigg)y'(z)+\bigg(\frac{\kappa_1 \kappa_2}{z(z-1)}-\frac{t (t-1) K}{z(z-1)(z-t)}\bigg)y(z)=0,
\end{equation}
where $\theta_0$, $\theta_1$ and $\theta_t$ are the characteristic exponents around the regular singular points $z_i=(z_0=0,z_1=1, z_t=t)$. The parameters $\theta_0$, $\theta_1$ and $\theta_t$ can be obtained by the Frobenius method around a given singular point. These are encoded in the Riemann symbol for the Heun equation:
\begin{equation}
    \mathcal{P}=\left(
\begin{array}{ccccc}
 0 & 1 & t & \infty  &   \\
 0 & 0 & 0 & \kappa _1 & z \\
 \theta _0 & \theta _1 & \theta _t & \kappa _2 &   \\
\end{array}
\right)=\left(
\begin{array}{ccccc}
 0 & 1 & \tau & \infty  &   \\
 0 & 0 & 0 & \alpha & z \\
 1-\gamma & 1-\delta & 1-\epsilon & \beta &   \\
\end{array}
\right).
\end{equation}
Comparing (\ref{eqStandardHeun}) and (\ref{eqHeunThetaForm}) one has:
\begin{equation}\label{eqHeuntoHeun}
    1-\theta _0=\gamma ,\quad 1-\theta _1=\delta ,\quad 1-\theta _t=\epsilon,\quad \kappa _1 \kappa _2=\alpha  \beta,\quad Kt(1-t)-t \kappa _1 \kappa _2=-k, \quad t=\tau,
\end{equation}
together with the following form of the Fuchs relation 
\begin{equation}
\theta_0+\theta_1+\theta_t+\kappa_1+\kappa_2=2.
\end{equation}
Obviously, if we substitute all expressions from (\ref{eqHeuntoHeun}) we restore the standard Fuchs relation given by (\ref{Fuch_rel}).

Finally, one can write the characteristic exponent around infinity in terms of $\kappa_1=\alpha$ and $\kappa_2=\beta$, namely $\kappa_1-\kappa_2=\alpha-\beta=\theta_\infty$.

\section{Gauss hypergeometric equation, Frobenius method and characteristic exponents}

The Gauss hypergeometric equation in canonical form is written by
\begin{equation}\label{EqGaus1}   
z(z-1) u''(z)+\big((a+b+1) z-c\big) u'(z)+a b u(x)=0.
\end{equation}
The characteristic exponents $\theta_{0,1,\infty}^{(hyp)}$ around the singular points ($0,1,\infty$) can be found by the Frobenius method. For example, around  $z=0$ we look for a solution of the form
\begin{equation}
    u(z)=(z-0)^r\sum\limits_{k=0}^{\infty} a_k (z-0)^k=z^r\sum\limits_{k=0}^{\infty} a_k z^k.
\end{equation}
where the roots for $r$, defining $\theta_0^{(hyp)}$ around  $z_0=0$,  can be obtained by the coefficient in front of the lowest power of $z$. The equation is called indicial equation. To find $r$ we first differentiate the solution with respect to $z$ once and twice, respectively:
\begin{align}
&u'(z)=\frac{d}{dz}\sum\limits_{k=0}^{\infty} a_k z^{(k+r)}=\sum\limits_{k=0}^{\infty} (k+r) a_k z^{(k+r-1)},
\\
&u''(z)=\frac{d^2}{dz^2}\sum\limits_{k=0}^{\infty} a_k z^{(k+r)}=\sum\limits_{k=0}^{\infty} (k+r)(k+r-1) a_k z^{(k+r-2)}.
\end{align}
Inserting these expressions in (\ref{EqGaus1}) one finds: 
\begin{align}
\sum\limits_{k=0}^{\infty} a_k z^{k+r}\big[(z-1) (k+r)(k+r-1)  z^{-1}+\big((a+b+1) z-c\big) (k+r) z^{-1}+a b \big]=0.
\end{align}
Here we extract the lowest power in $z$:
\begin{equation}
    a_0 r (1-c-r) z^{r-1}+\sum\limits_{k=1}^{\infty} a_k z^{k+r}\big[(z-1) (k+r)(k+r-1)  z^{-1}+\big((a b+1) z-c\big) (k+r) z^{-1}+a b \big]=0.
\end{equation}
The coefficient in front of $z^{r-1}$ characterizes the indicial equation around $z_0=0$. Its solutions give the characteristic exponent $\theta_0$ around $z_0=0$:
\begin{equation}
 a_0 r (1-c -r)=0\quad \Rightarrow \quad r_1=0,\quad r_2=1-c, \quad a_0\neq 0.
\end{equation}

Similar calculations can be made for the other two singularities $z_1=1$ and $z_\infty=\infty$, which has $r=(0,c-a-b)$ and $r=(a,b)$. Therefore, the Riemann symbol for the Gauss hypergeometric equation can be written by
\begin{equation}
    \mathcal{P}
=\left(
\begin{array}{cccc}
 0 & 1 &  \infty  &   \\
 0 & 0  & a & z \\
 1-c & c-a-b & b &   \\
\end{array}
\right)=\left(
\begin{array}{cccc}
 0 & 1 &  \infty  &   \\
 0 & 0  & a & z \\
 \theta_0^{(hyp)} & \theta_1^{(hyp)} & b &   \\
\end{array}
\right).
\end{equation}

Finally, this leads to the following characteristic exponents:
\begin{equation}\label{eqCharExpHyp}
    \theta_0^{(hyp)}=1-c,\quad \theta_1^{(hyp)}=c-a-b,\quad \theta_\infty^{(hyp)}=a-b.
\end{equation}

\section{Painlev\'e equations} \label{appPainleve}
The Painlev\'e equations are a set of non-linear integrable differential equations. In the context of this work, they appear as monodromy preserving flows of Fuchsian differential equations which posses the Painlev\'e property, meaning that the only movable singularities are simple poles. Another important property of these equations is that they are part of a coalescence cascade which allows one to transform a higher order Painlev\'e equation to one of lower order satisfied by the same $\tau$-function. This is due to the fact that the locations of the singular points are arbitrary and can be changed with a transformation, as explained in Section \ref{isomTh}. Thus, upon swapping the positions of two singularities the monodromy data remains the same. This means that the equations must be invariant under the reflection group of four elements and this results in the possible transformation between equations. A simplified coalescence diagram with the corresponding original differential equations can be found below:
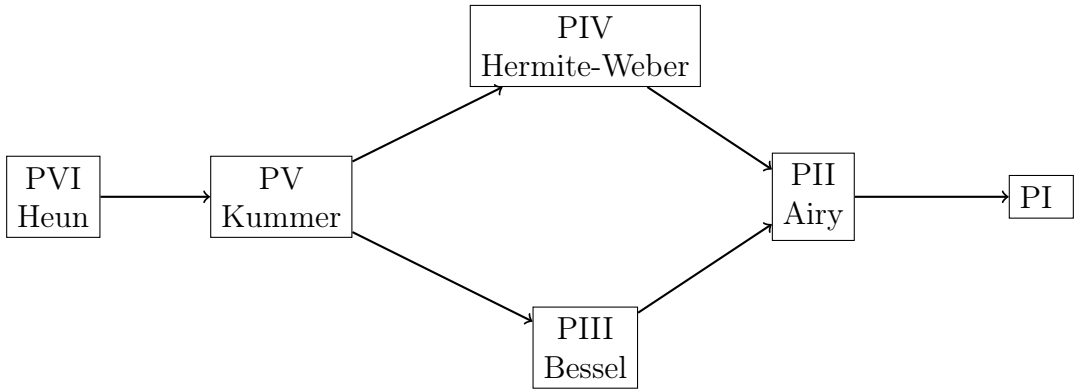
\begin{figure}[H]
    \centering
    \begin{tikzpicture}
        \node[draw, align=center] (PVI) at (0, 0)    { PVI  \\ Heun};
        \node[draw, align=center] (PV) at  (3, 0)    { PV   \\ Kummer};
        \node[draw, align=center] (PIV) at  (7, 2)   { PIV  \\ Hermite-Weber};
        \node[draw, align=center] (PIII) at  (7, -2) { PIII  \\ Bessel};
        \node[draw, align=center] (PII) at  (10, 0)  { PII  \\ Airy};
        \node[draw, align=center] (PI) at  (13, 0) { PI };

        \draw[thick, ->] (PVI) -- (PV);
        \draw[thick, ->] (PV) -- (PIII);
        \draw[thick, ->] (PV) -- (PIV);
        \draw[thick, ->] (PIII) -- (PII);
        \draw[thick, ->] (PIV) -- (PII);
        \draw[thick, ->] (PII) -- (PI);  
    \end{tikzpicture}
    \caption{An incomplete diagram showing the possible transformations between Painlev\'e equations. It is important to note that some of the equations (in particular PV and PIII) have degenerate versions, which leads to many more nodes in the diagram. More detailed diagram  can be found in e.g. \cite{chekhov:2017} For exact form of the transformation. }
    \label{painDeg}
\end{figure}
The explicit form of the six nonlinear Painlev\'e equations is given by:
\begin{equation}
 \text{PI}:\quad \lambda'' = 6 \lambda^2 + t,
\end{equation}
\begin{equation}
 \text{PII}:\quad \lambda''=2\lambda^3 + t \lambda + \alpha,
\end{equation}
\begin{equation}
 \text{PIII}:\quad t\lambda \lambda'' = t \lambda'^2 - \lambda \lambda' + \delta t + \beta \lambda + \alpha \frac{\lambda^3}{t} + \gamma \frac{\lambda^4}{t},
\end{equation}
\begin{equation}
 \text{PIV}:\quad \lambda \lambda'' = \frac{1}{2} \lambda'^2 + \beta + 2 (t^2 - \alpha)\lambda^2 + 4 t \lambda^3 + \frac{3}{2} \lambda^4,
\end{equation}
\begin{equation}
 \text{PV}:\quad \lambda'' = \bigg(\frac{1}{2\lambda} + \frac{1}{\lambda-1}\bigg) \lambda'^2 - \frac{1}{t} \lambda' + \frac{ (\lambda - 1 )^2}{t^2} \bigg(\alpha \lambda + \frac{\beta}{\lambda} \bigg) + \gamma \frac{\lambda}{t} + \delta \frac{\lambda (\lambda +1)}{\lambda-1},
\end{equation}
\begin{align}
   \text{PVI}:\quad \lambda'' &= \frac{1}{2} \bigg(\frac{1}{\lambda} + \frac{1}{\lambda-1} + \frac{1}{\lambda-t}\bigg) {\lambda'}^2 - 
    \bigg( \frac{1}{t} + \frac{1}{t-1} + \frac{1}{\lambda-t}\bigg) \lambda' 
    \\
    &+ 
    \frac{\lambda(\lambda -1)(\lambda -t)}{t^2(t-1)^2} \bigg(\alpha + \beta \frac{t}{\lambda^2} +
    \gamma \frac{t-1}{(\lambda -1)^2}+\delta \frac{t(t-1)}{(\lambda - t)^2} \bigg).
\end{align}

\section{Casimir operator for $SU(2)$}\label{appB}

Here we explicitly show how to find the Casimir operator of the $SU(2)$ group. Let us start by the parametrization of the $SU(2)$ group element given by
\begin{align}
&U(\varphi,\theta,\psi)  = U_z(\varphi)U_y(\theta)U_x(\psi)= e^{i\sigma_3\frac{\varphi}{2}} e^{i\sigma_2\frac{\theta}{2}} e^{i\sigma_3\frac{\psi}{2}} \nonumber\\[5pt]
 & = \begin{pmatrix}  e^{i\frac{\varphi}{2}} & 0 \\[5pt] 0 & e^{-i\frac{\varphi}{2}}\end{pmatrix}
 \begin{pmatrix} \cos\frac{\theta}{2} & \sin\frac{\theta}{2} \\[5pt] -\sin\frac{\theta}{2} & \cos\frac{\theta}{2}\end{pmatrix}
 \begin{pmatrix}  e^{i\frac{\psi}{2}} & 0 \\[5pt] 0 & e^{-i\frac{\psi}{2}}\end{pmatrix} = \begin{pmatrix} \cos\frac{\theta}{2}e^{\frac{i}{2}(\varphi+\psi)} & \sin\frac{\theta}{2} e^{-\frac{i}{2}(-\varphi+\psi)} \\[5pt] -\sin\frac{\theta}{2} e^{\frac{i}{2}(-\varphi+\psi)} & \cos\frac{\theta}{2}e^{-\frac{i}{2}(\varphi+\psi)}\end{pmatrix}.
\end{align}
The Euler angles $\theta,\varphi$ and $\psi$ take values within the intervals $0 \leq \theta \leq \pi,\, \theta\leq \varphi \leq 2\pi$ and $0\leq \psi \leq 4\pi$. The left and right Maurer–Cartan one-forms (note that $dU^{-1}U = -U^{-1}dU$) are given by 
\begin{equation}
	R=U^{-1}dU=\frac{i}{2}\sigma_kR_k, \qquad L=dU\,U^{-1}=\frac{i}{2}\sigma_kL_k.
\end{equation}
The components of the Maurer–Cartan forms in the basis given by the
Pauli matrices are written as
\begin{align}
	& R_1=-\sin\psi d\theta+\cos\psi\sin\theta d\phi, & & L_1=\sin\phi d\theta -\cos\phi \sin\theta d\psi, \nonumber \\
	& R_2=\cos\psi d\theta + \sin\psi\sin\theta d\theta, & & L_2=\cos\phi d\theta +\sin\phi\sin\theta d\psi, \nonumber \\
	& R_3= d\psi + \cos\theta d\phi, & & L_3=d\phi +\cos\theta d\psi.
	\label{A.4}
\end{align}
Clearly, they satisfy the Maurer–Cartan equations
\begin{equation}
	dR_n=\frac{1}{2}\varepsilon_{nmk}R_m\wedge R_k, \qquad dL_n=-\frac{1}{2}\varepsilon_{nmk} L_m\wedge L_k.
\end{equation}

The left and right forms on the group $SU(2)$ are dual to the vector ﬁeld $\xi_k$, components of which form the standard basis of the Lie algebra on the group $SU(2)$:
\begin{equation}
	\langle \xi_k^{(R)},R_m\rangle =\delta_{km}, \qquad \langle \xi_k^{(L)},L_m\rangle =\delta_{mk}.
\end{equation}
Here the right and left Killing vectors are related with generators of rotations about the corresponding axis of Cartesian coordinates. They can be written in terms of the Euler parameterization as
\begin{align}
	& \xi_1^{(R)}= -\cot\theta\cos\psi\frac{\partial}{\partial\psi} -\sin\psi \frac{\partial}{\partial\theta} +\frac{\cos\psi}{\sin\theta}\frac{\partial}{\partial\phi}, \nonumber \\
	& \xi_2^{(R)}= -\cot\theta\sin\psi\frac{\partial}{\partial\psi} +\cos\psi \frac{\partial}{\partial\theta} +\frac{\sin\psi}{\sin\theta}\frac{\partial}{\partial\phi}, \nonumber \\
	& \xi_3^{(R)}= \frac{\partial}{\partial\psi}
	\label{killing-su2},
\end{align}
and
\begin{align}
	& \xi_1^{(L)}= -\frac{\cos\phi}{\sin\theta}\frac{\partial}{\partial\psi} + \sin\phi \frac{\partial}{\partial\theta} +\cot\theta\cos\phi\frac{\partial}{\partial\phi}, \nonumber \\
	& L_2^{(L)}= \frac{\sin\phi}{\sin\theta}\frac{\partial}{\partial\psi} + \cos\phi \frac{\partial}{\partial\theta} -\cot\theta\sin\phi\frac{\partial}{\partial\phi}, \nonumber \\
	& L_3^{(L)}=\frac{\partial}{\partial\phi}.
\end{align}

Note that the generators of the left and right rotations commute, while left and right Killing vectors satisfy the $SU(2)$ Lie algebra
\begin{equation}
	[\xi_m^{(R)},\xi_n^{(R)}]=-\varepsilon_{mnk}\,\xi_k^{(R)}, \qquad [\xi_m^{(R)},\xi_n^{(R)}]=\varepsilon_{mnk}\,\xi_k^{(R)}, \qquad [\xi_m^{(R)},\xi_n^{(R)}]=0.
\end{equation}

 The vector ﬁelds on the sphere $S^3$ are related with the angular momentum operator as
\begin{equation}
	L_n^{(R)}=-i\xi_n^{(R)}, \qquad L_n^{(L)}=i\xi_n^{(L)}.
\end{equation}
It follows from the above relation that the components of the operator of angular momentum satisfy the usual commutation relation, which does not distinguish between left and right rotations:
\begin{equation}
	[L_n,L_m]=i\varepsilon_{nmk}L_k.
\end{equation}

The quadratic Casimir operator/matrix related to the Pauli basis is:
\begin{equation}
	\bm{C}^2=L_1^2+L_2^2+L_3^2,
\end{equation}
which explicitly yields
\begin{multline}
	\bm{C}^2= \frac{\partial^2}{\partial\theta^2} +\cot\theta \frac{\partial}{\partial\theta} + \frac{1}{\sin^2\theta}\frac{\partial^2}{\partial\phi^2} + 2\frac{\cot\theta}{\sin\theta} \frac{\partial}{\partial\phi}\frac{\partial}{\partial\psi} + \cot^2\theta\frac{\partial^2}{\partial\psi^2}+ \left(\frac{\partial}{\partial \psi} \right)^2\\
	= \frac{1}{\sin \theta} \frac{\partial}{\partial \theta} \sin \theta
	\frac{\partial}{\partial \theta} + \frac{1}{\sin^2 \theta} \left(\frac{\partial}{\partial \phi} + \cos
	\theta \frac{\partial}{\partial \psi}  \right)^2 + \left(\frac{\partial}{\partial \psi} \right)^2.
\end{multline}

\bibliographystyle{ieeetr}
\bibliography{References}
%%%%%%%%%%%%%%%%%%%%%
\end{document}